\documentstyle[12pt,epsf]{article}
\renewcommand{\bar}[1]{\overline{#1}}

\begin{document}

\begin{flushright}
USM-TH-95
\\ CPT-2000/P.4032
\end{flushright}
\bigskip\bigskip

\centerline{\Large \bf The Flavor and Spin Structure of Hyperons}

\centerline{\Large \bf from Quark Fragmentation}

\vspace{22pt}
\centerline{\bf
Bo-Qiang Ma\footnote{e-mail: mabq@phy.pku.edu.cn}$^{a}$,
Ivan Schmidt\footnote{e-mail: ischmidt@fis.utfsm.cl}$^{b}$,
Jacques Soffer\footnote{e-mail: Jacques.Soffer@cpt.univ-mrs.fr}$^{c}$,
Jian-Jun Yang\footnote{e-mail: jjyang@fis.utfsm.cl}$^{b,d}$}

\vspace{8pt}

{\centerline {$^{a}$Department of Physics, Peking University,
Beijing 100871, China\footnote{Mailing address},}}

{\centerline {CCAST (World Laboratory),
P.O.~Box 8730, Beijing 100080, China,}}

{\centerline {and Institute of Theoretical Physics, Academia
Sinica, Beijing 100080, China}

{\centerline {$^{b}$Departamento de F\'\i sica, Universidad
T\'ecnica Federico Santa Mar\'\i a,}}

{\centerline {Casilla 110-V, 
Valpara\'\i so, Chile}

{\centerline {$^{c}$Centre de Physique Th$\acute{\rm{e}}$orique,
CNRS, Luminy Case 907,}}

{\centerline { F-13288 Marseille Cedex 9, France}}

{\centerline {$^{d}$Department of Physics, Nanjing Normal
University,}}

{\centerline {Nanjing 210097, China}}

\vspace{10pt}
\begin{center} {\large \bf Abstract}

\end{center}
We systematically study the hadron longitudinal polarizations of
the octet baryons at large $z$ from quark fragmentations in
$e^+e^-$-annihilation, polarized charged lepton deep inelastic
scattering (DIS) process, and neutrino (antineutrino) DIS process,
based on predictions of quark distributions for the octet baryons
in the SU(6) quark-spectator-diquark model and a perturbative QCD
based counting rule analysis. We show that the
$e^+e^-$-annihilation and polarized charged lepton DIS process are
able to distinguish between the two different predictions of the
hyperon polarizations. We also find that the neutrino/antineutrino
DIS process is ideal in order to study both the valence content of
the hyperons and the antiquark to hyperon (quark to anti-hyperon)
fragmentations, which might be related to the sea content of
hyperons.

\vfill
\centerline{PACS numbers: 14.20.Jn, 13.65.i, 13.87.Fh,
13.88.+e}

\vfill
\centerline{To be published in Phys. Rev. D}
\vfill

\newpage
\section{Introduction}

Recently there have been some significant progress in
understanding the flavor and spin structure of the
$\Lambda$-hyperon from various fragmentation processes, both
theoretically [1-18] and experimentally [19-24]. One of the most
interesting new observations is related to the polarizations of
the up ($u$) and down ($d$) quarks inside the $\Lambda$. In the
naive quark model, the $\Lambda$ spin is totally provided by the
strange ($s$) quark, and the $u$ and $d$ quarks are unpolarized.
Based on novel results concerning the proton spin structure from
deep inelastic scattering (DIS) experiments and SU(3) symmetry
between the octet baryons, it was found that the $u$ and $d$
quarks of the $\Lambda$ should be negatively polarized
\cite{Bur93}. It was also pointed out that the $u$ and $d$
polarizations in the $\Lambda$ are related to the $s$
polarizations of the proton \cite{Ma99}. However, based on a
perturbative QCD (pQCD) counting rule analysis
\cite{countingr,Bro95} and an SU(6) quark-spectator-diquark model
\cite{Ma96}, it was later predicted \cite{MSY2,MSY3,MSSY5} that,
although the $u$ and $d$ quarks of the $\Lambda$ might be
unpolarized or negatively polarized in the integrated Bjorken
range $0 \leq x \leq 1$, they should be positively polarized at
large $x$. This prediction seems to be supported by all available
data from longitudinally polarized $\Lambda$ fragmentations in
$e^+e^-$-annihilation \cite{ALEPH96,DELPHI95,OPAL97}, polarized
charged lepton DIS process \cite{HERMES,E665}, and most recently,
neutrino (antineutrino) DIS process \cite{NOMAD}.

However, there are still many unknowns to be explored before we
can arrive at some definite conclusion on the $\Lambda$ quark and spin
structure. First, what one actually measures in experiments are the
hyperons from quark fragmentation, and therefore one needs a
relation between  quark distributions and fragmentation
functions. The Gribov-Lipatov relation \cite{GLR,Bro97}
\begin{equation}
D_q^h(z) \sim z \, q_h(z)~, \label{GLR}
\end{equation}
with $D_q^h(z)$ being the fragmentation function for a quark $q$
splitting into a hadron $h$ with longitudinal momentum fraction
$z$, and $q_h(z)$ being the quark distribution of finding the
quark $q$ inside the hadron $h$ carrying a momentum fraction
$x=z$, was used to connect the fragmentation functions from
predictions on the quark distributions
\cite{MSY2,MSY3,MSY4,MSSY5,MSSY6,Bro97}. However, such a relation
is only known to be valid near $z \to 1$ and on a certain energy
scale $Q^2_0$ in leading order approximation, and there are
serious doubts, coming from both theory and experiment, as to
whether this relation can be applied anywhere else. Thus it might
be more practical to consider Eq.~(\ref{GLR}) as a
phenomenological Ansatz to parameterize the quark to $\Lambda$
fragmentation functions, and then check the validity and
reasonableness of the method by comparing the theoretical
predictions with the experimental data. In fact, most other
theoretical estimates \cite{Bur93,Kot98,Bor98,Flo98b,Ash99,Liu00}
on the quark fragmentation functions are also based on some
knowledge of quark distributions. Second, there are still
uncertainties on the quark distributions, even for the valence
components. For example, the flavor structure of the $\Lambda$
differs significantly in the pQCD based analysis and the
quark-diquark model: the ratio $u(x)/s(x)$ at $x=1$ is 1/2 in the
pQCD analysis whereas it is 0 in the quark-diquark model
\cite{MSY2}. But the two models have similar predictions of the
$\Lambda$ polarizations of fragmentations in $e^+e^-$-annihilation
\cite{MSY3}, polarized charged lepton DIS process
\cite{MSY2,MSSY5}, and neutrino (antineutrino) process
\cite{MSSY5}, and it is still difficult to distinguish between the
two different model predictions with the available data. Therefore
we still need to look for new quantities and kinematic regions
where the distinction of the different predictions is feasible.

It has been pointed out \cite{MSY4} that the two different
predictions of the quark distributions for the $\Sigma^{\pm}$ and
$\Xi^-$ hyperons can be directly tested in Drell-Yan processes of
charged hyperon beams on the nucleon target. However, it might
take a long time for performing such experiments, since the
technique on the charged hyperon beams still needs improvement for
precision experimental purposes. By comparison, the detection
technique of $\Sigma$ and $\Xi$ hyperons is more mature in order
to measure the various quark to hyperon fragmentation functions
\cite{SigmaP,XiP,Sigma0P}. Except for the $\Sigma^0$, which decays
electromagnetically, all other hyperons in the octet baryons have
their major decay modes mediated by the weak interaction. Because
these weak decays do not conserve parity, information from their
decay products can be used to determine their polarization
\cite{SigmaP,XiP}. The polarization of $\Sigma^0$ can be also
re-constructed from the dominant decay chain $\Sigma^0 \to \Lambda
\gamma$ and $\Lambda \to p \pi^-$ \cite{Sigma0P}. Therefore we can
use the measurable fragmentation functions to extract information
on the spin and flavor content of hyperons with the available
experimental facilities. From another point of view, studying the
fragmentation functions of various hyperons is also interesting in
itself, in addition to its connection to the quark distributions.
The purpose of this paper is to study the longitudinal
polarizations of various hyperon fragmentations in
$e^+e^-$-annihilation, charged lepton DIS process, and neutrino
DIS process, based on predictions of quark distributions for the
hyperons in the pQCD analysis and quark-diquark model. This is
useful as a systematically survey of various hyperon fragmentation
functions, as well as checking different model predictions.

The paper is organized as follows. In Sec.~II we will present a
brief review of the quark distributions for the octet baryons in
both the quark-diquark model and the pQCD based analysis. For the
pQCD based analysis we make a new set of leading order quark
distributions for the valence quarks with SU(3) symmetry between
the octet baryons. This set of quark distributions has no other
free parameters, therefore it has predictive ability. In Sec.~III
we calculate the baryon polarizations in $e^+e^-$-annihilation for
the octet baryons at two energies: LEP I at the $Z$ resonance
$\sqrt{s} \approx 91$~GeV and LEP II at $\sqrt{s} \approx
200$~GeV. We find that predicted polarizations for $\Sigma$'s are
quite different in the two models in the medium to large $z$
region, and a distinction between different predictions can be
checked by measuring the $\Sigma^{\pm}$ polarizations in
$e^+e^-$-annihilation. Sec.~IV contains our predictions for the
baryon and anti-baryon polarizations of the octet baryons in
polarized charged lepton DIS process. We find that the $\Xi^0$ has
the biggest difference for the spin transfer in the two different
models, and we propose to measure the $\Xi^0$ polarization in this
process as a sensitive test of different predictions. Sec.~V is
devoted to the baryon and anti-baryon polarizations of the octet
baryons in neutrino (antineutrino) DIS process. We find that the
neutrino (antineutrino) DIS process is ideal to test different
predictions concerning both the valence and sea content of the
hyperons. Finally, we present a summary of our new results
together with our conclusions in Sec.~VI.

\section{The quark flavor and spin structure of octet baryons}

The valence quark distributions of the octet baryons, in both the
light-cone SU(6) quark-spectator-diquark model \cite{Ma96} and the
pQCD based counting rule analysis \cite{Bro95}, have been
discussed in previous publications \cite{MSY2,MSY3,MSY4}. Here we
briefly outline the main ingredients and new features that are
present in this paper for the later applications.

\subsection{The light-cone SU(6) quark-spectator-diquark model}

The application of the quark-spectator-diquark model to discuss
the quark distributions of nucleons at large $x$ can be traced
back to a work of Feynman, to explain the unexpected behavior of
$F_2^n(x)/F_2^p(x)=1/4$ at $x \to 1$ in the experimental
observation at that time \cite{Fey72}. There have been many
developments along this line \cite{DQM}, and the light-cone SU(6)
quark-spectator-diquark model \cite{Ma96} is a revised version
with the new ingredient of Wigner-Melosh rotation effect
\cite{Ma91b,Ma98} taken into account. This model does not
necessarily break the bulk SU(6) symmetry of the wavefunction, and
is successful in describing the large $x$ behavior of polarized
structure functions of nucleons. The main idea of this model is to
start from the three quark  SU(6) quark model wavefunction of the
baryon and then if any one of the quarks is probed, reorganize
the other two quarks in terms of two quark wavefunctions with
spins 0 or 1 (scalar and vector diquarks), i.e., the diquark
serves as an effective particle which is called the spectator. The
advantage of this model is that the non-perturbative effects such
as gluon exchanges between the two spectator quarks or other
non-perturbative gluon effects in the hadronic debris can be
effectively taken into account by the mass {\it et al.} of the
diquark spectator. So the complicated many-particle system can be
effectively treated by a simple two particle system technique
\cite{BHL,Hua94}. The mass difference between the scalar and
vector diquarks is proved to be important for producing
consistency with experimental observations \cite{Ma96}, in
comparison with the naive quark model with exact SU(6) symmetry.

The light-cone SU(6) quark-spectator-diquark model \cite{Ma96} is
extended to the $\Lambda$-hyperon in Refs.~\cite{MSY2,MSY3}. It is
interesting to notice that the mass difference between the scalar
and vector diquarks causes a suppression of  anti-parallel spin
components of quark distributions at large $x$, and as a
consequence the totally non-polarized $u$ and $d$ quarks should be
positively polarized at large $x$. Surprisingly, the predictions
of the model with naive parameters, and without any adjustment,
have been proved to be successful in describing all of the
available data of $\Lambda$ polarizations in $e^+e^-$-annihilation
\cite{MSY3}, polarized charged lepton DIS process
\cite{MSY2,MSSY5}, and neutrino (antineutrino) DIS process
\cite{MSSY5}, by adopting the simple Ansatz of the Gribov-Lipatov
relation Eq.~(\ref{GLR}), in order to connect fragmentation
functions with distribution functions. It is natural that we
should try to check or refine the validity of this method by
exploring hadron fragmentations of other octet baryons. The
extension of the light-cone SU(6) quark-spectator-diquark model to
the octet baryons has been done in Ref.~\cite{MSY4}, and we
outline the main ingredients in the following.

The unpolarized quark distribution for a quark with flavor $q$
inside a hadron $h$ is expressed as
\begin{equation}
q(x)= c_q^S a_S(x) + c_q^V a_V(x),
\end{equation}
where $c_q^S$ and $c_q^V$ are the weight coefficients determined
by the SU(6) quark-diquark model wavefunctions and are different
for various baryons, and $a_D(x)$ ($D=S$ for scalar spectator or
$V$ for axial vector spectator) can be expressed  in terms of the
light-cone momentum space wavefunction $\varphi (x, {\mathbf
k}_\perp)$ as
\begin{equation}
a_{D}(x) \propto  \int [\rm{d}^2 {\mathbf k}_\perp] |\varphi (x,
{\mathbf k}_\perp)|^2 \hspace{1cm} (D=S \hspace{0.2cm} or
\hspace{0.2cm} V)
\end{equation}
which is normalized such that $\int_0^1 {\mathrm d} x a_D(x)=3$
and denotes the amplitude for quark $q$ to be scattered while the
spectator is in the diquark state $D$. We employ the
Brodsky-Huang-Lepage (BHL) prescription \cite{BHL} of the
light-cone momentum space wavefunction for the quark-diquark
\begin{equation}
\varphi (x, {\mathbf k}_\perp) = A_D \exp \{-\frac{1}{8\alpha_D^2}
[\frac{m_q^2+{\mathbf k}_\perp ^2}{x} + \frac{m_D^2+{\mathbf
k}_\perp^2}{1-x}]\},
\end{equation}
with the parameter $\alpha_D=330$~MeV. Other parameters such as
the quark mass $m_q$, vector(scalar) diquark mass $m_{D}$
($D=S,V$) for the octet baryons are listed in Table 1.

One needs to introduce the Melosh-Wigner correction factor
\cite{Ma91b,Ma98} in order to calculate the polarized quark
distributions
\begin{equation}
\Delta q(x)= \tilde{c}_q^S \tilde{a}_S(x) + \tilde{c}_q^V
\tilde{a}_V(x),
\end{equation}
where the coefficients $\tilde{c}_q^S$ and  $\tilde{c}_q^V$ are
also determined by the SU(6) quark-diquark wavefunctions, and
$\tilde{a}_D(x)$ is expressed as
\begin{equation}
\tilde{a}_{D}(x) = \int [\rm{d}^2 {\mathbf k}_\perp]
W_D(x,{\mathbf k}_\perp) |\varphi (x, {\mathbf k}_\perp)|^2
\hspace{1cm} (D=S \hspace{0.2cm} or \hspace{0.2cm} V)
\end{equation}
where
\begin{equation}
W_D(x,{\mathbf k}_{\perp}) =\frac{(k^+
+m_q)^2-{\mathbf k}^2_{\perp}} {(k^+ +m_q)^2+{\mathbf
k}^2_{\perp}} \label{eqM1},
\end{equation}
with $k^+=x {\cal M}$ and ${\cal M}^2=\frac{m^2_q+{\mathbf
k}^2_{\perp}}{x}+\frac{m^2_D+{\mathbf k}^2_{\perp}}{1-x}$. The
weight coefficients for all octet baryons,  $c_q^S$, $c_q^V$,
$\tilde{c}_q^S$, and  $\tilde{c}_q^V$, can be also found in Table
1. We thus have all the formalism to calculate the quark
distributions of the octet baryons. We should mention that the SU(3)
symmetry between the octet baryons is in principle maintained, but
the mass difference between different quarks and diquarks breaks
the SU(3) symmetry explicitly. There is still degrees of freedom
to refine the model by improving the parameters and the explicit
forms of the momentum space wavefunctions.

\vspace{0.5cm}
\newpage

\centerline{Table~1~~ The quark distribution functions of octet
baryons in SU(6) quark-diquark model}

\vspace{0.3cm}

\begin{footnotesize}
\begin{center}
\begin{tabular}{|c||c|c||c|c||c|c|c|}\hline
 ~~~~~~~~~~~~~~~& $~~~~~~~$ & ~~~~~~~~~~~~~~~& $~~~~~~$ &~~~~~~~~~~~~~~~&
$m_q$ & $m_V$ & $m_S$
\\
Baryon& $~~q ~~$ & ~~~~~~~~~~~~~~~& $~~\Delta q ~~$
&~~~~~~~~~~~~~~~& (MeV) & (MeV) & (MeV)
\\ \hline
 ~~~~p~~~~ & $~~u~~$ &$\frac{1}{6}a_V+\frac{1}{2}a_S $ &
 $\Delta u$ & -$\frac{1}{18}\tilde{a}_V+\frac{1}{2}\tilde{a}_S $ &
330 & 800 & 600 \\ \cline{2-8}
 (uud)& $~~d~~$ &$\frac{1}{3}a_V$ &
 $\Delta d$ & -$\frac{1}{9}\tilde{a}_V$ &
330 & 800 & 600 \\ \cline{1-8}
 ~~~~n~~~~ & $~~u~~$ &$\frac{1}{3}a_V $ &
 $\Delta u$ & -$\frac{1}{9}\tilde{a}_V $ &
330 & 800 & 600 \\ \cline{2-8}
 (udd)& $~~d~~$ &$\frac{1}{6}a_V+\frac{1}{2} a_S$ &
 $\Delta d$ & -$\frac{1}{18}\tilde{a}_V+\frac{1}{2}\tilde{a}_S$ &
330 & 800 & 600 \\ \cline{1-8}
 $~~~~\Sigma^{+}~~~~$ & $~~u~~$ &$\frac{1}{6}a_V+\frac{1}{2}a_S $ &
 $\Delta u$ & -$\frac{1}{18}\tilde{a}_V+\frac{1}{2}\tilde{a}_S $ &
330 & 950 & 750 \\ \cline{2-8}
 (uus)& $~~s~~$ &$\frac{1}{3}a_V$ &
 $\Delta s$ & -$\frac{1}{9}\tilde{a}_V$ &
480 & 800 & 600 \\ \cline{1-8}
 $~~~~\Sigma^{0}~~~~$ & $~~u~~$ &$\frac{1}{12}a_V+\frac{1}{4}a_S $ &
 $\Delta u$ & -$\frac{1}{36}\tilde{a}_V+\frac{1}{4}\tilde{a}_S $ &
330 & 950 & 750 \\ \cline{2-8}
 (uds)& $~~d~~$ & $\frac{1}{12}a_V+\frac{1}{4}a_S $ &
 $\Delta d$ & -$\frac{1}{36}\tilde{a}_V+\frac{1}{4}\tilde{a}_S $ &
330 & 950 & 750 \\ \cline{2-8}
 $~~~~$ & $~~s~~$ & $\frac{1}{3}a_V$ &
 $\Delta s$ & -$\frac{1}{9}\tilde{a}_V $ &
480 & 800 & 600 \\ \cline{1-8} $~~~~\Sigma^{-}~~~~$ & $~~d~~$
&$\frac{1}{6}a_V+\frac{1}{2}a_S $ &
 $\Delta d$ & -$\frac{1}{18}\tilde{a}_V+\frac{1}{2}\tilde{a}_S $ &
330 & 950 & 750 \\ \cline{2-8}
 (dds)& $~~s~~$ &$\frac{1}{3}a_V$ &
 $\Delta s$ & -$\frac{1}{9}\tilde{a}_V$ &
480 & 800 & 600 \\ \cline{1-8} $~~~~\Lambda^{0}~~~~$ & $~~u~~$
&$\frac{1}{4}a_V+\frac{1}{12}a_S $ &
 $\Delta u$ & -$\frac{1}{12}\tilde{a}_V+\frac{1}{12}\tilde{a}_S $ &
330 & 950 & 750 \\ \cline{2-8}
 (uds)& $~~d~~$ & $\frac{1}{4}a_V+\frac{1}{12}a_S $ &
 $\Delta d$ & -$\frac{1}{12}\tilde{a}_V+\frac{1}{12}\tilde{a}_S $ &
330 & 950 & 750 \\ \cline{2-8}
 $~~~~$ & $~~s~~$ & $\frac{1}{3}a_S$ &
 $\Delta s$ & $\frac{1}{3}\tilde{a}_S $ &
480 & 800 & 600 \\ \cline{1-8}
 $~~~~\Xi^{-}~~~~$ & $~~d~~$ &$\frac{1}{3}a_V $ &
 $\Delta d$ & -$\frac{1}{9}\tilde{a}_V $ &
330 & 1100 & 900 \\ \cline{2-8}
 (dss)& $~~s~~$ &$\frac{1}{6}a_V+\frac{1}{2} a_S$ &
 $\Delta s$ & -$\frac{1}{18}\tilde{a}_V+\frac{1}{2}\tilde{a}_S$ &
480 & 950 & 750 \\ \cline{1-8}
 $~~~~\Xi^{0}~~~~$ & $~~u~~$ &$\frac{1}{3}a_V $ &
 $\Delta u$ & -$\frac{1}{9}\tilde{a}_V $ &
330 & 1100 & 900 \\ \cline{2-8}
 (uss)& $~~s~~$ &$\frac{1}{6}a_V+\frac{1}{2} a_S$ &
 $\Delta s$ & -$\frac{1}{18}\tilde{a}_V+\frac{1}{2}\tilde{a}_S$ &
480 & 950 & 750 \\ \cline{1-8}
\end{tabular}
\end{center}
\end{footnotesize}

\vspace{0.5cm}

\subsection{The perturbative QCD counting rule analysis}

We now look at the pQCD counting rule analysis of the quark
distributions based on minimally connected tree graphs of hard
gluon exchanges. In the region $x \to 1$ such an approach can give
rigorous predictions for the behavior of distribution functions
\cite{Bro95}. In particular, it predicts ``helicity retention'',
which means that the helicity of a valence quark will match that
of the parent nucleon. Explicitly, the quark distributions of a
hadron $h$ have been shown to satisfy the counting rule
\cite{countingr},
\begin{equation}
q_h(x) \sim (1-x)^p, \label{pl}
\end{equation}
where
\begin{equation}
p=2 n-1 +2 \Delta S_z.
\end{equation}
Here $n$ is the minimal number of the spectator quarks, and
$\Delta S_z=|S_z^q-S_z^h|=0$ or $1$ for parallel or anti-parallel
quark and hadron helicities, respectively \cite{Bro95}. Therefore
the anti-parallel helicity quark distributions are suppressed by a
relative factor $(1-x)^2$, and consequently $\Delta q(x)/q(x) \to
1$ as $x \to 1$. A further input of the model, explained in detail
in Ref.~\cite{Bro95}, is to retain the SU(6) ratios only for the
parallel helicity distributions at large $x$, since in this region
SU(6) is broken into SU(3)$^{\uparrow}\times$SU(3)$^{\downarrow}$.
With such power-law behaviors of quark distributions, the ratio
$d(x)/u(x)$ of the nucleon was predicted \cite{Far75} to be 1/5 as
$x \to 1$, and this gives $F_2^n(x)/F_2^p(x)=3/7$, which is
(comparatively) close to the quark-diquark model result $1/4$
\cite{Fey72,DQM}. From the different power-law behaviors for
parallel and anti-parallel quarks, one easily finds that $\Delta
q/q =1$ as $x \to 1$ for any quark with flavor $q$, unless the $q$
quark is completely negatively polarized \cite{Bro95}. This
prediction is quite different from the quark-diquark model
prediction that $\Delta d(x)/d(x)=-1/3$ as $x \to 1$ for the
nucleon \cite{Ma96}. The most recent analysis \cite{Yang99,Sch00}
of experimental data for several processes seems to support the
pQCD based prediction of the unpolarized quark behaviors
$d(x)/u(x)=1/5$ at $x \to 1$, but there is still no definite test
of the polarized quark behaviors $\Delta d(x)/d(x)$ since the $d$
quark is the non-dominant quark for the proton and does not play a
dominant role at large $x$.

The extension of pQCD based counting rule analysis to the
$\Lambda$ is done in Refs.~\cite{MSY2,MSY3}, where it is shown
that the ratio $u(x)/s(x) \to 1/2$ at $x \to 1$, and this is
different from the quark-diquark model prediction that $u(x)/s(x)
\to 0$. However, the pQCD based analysis also predicts the
positively polarized $u$ and $d$ quarks at large $x$ and this is
similar to the quark-diquark model prediction. It is interesting
that with some adjustment to the parameterizations, the pQCD
analysis can also reproduce all available data of $\Lambda$
polarizations in fragmentations \cite{MSY2,MSY3,MSSY5}. The pQCD
analysis of quark distributions has also been extended to the
octet baryons \cite{MSY4}, and predictions on the Drell-Yan
proceses involving hyperons have been given. This paper is
purposed to study the quark distributions of all octet hyperons
through fragmentations, similar to the $\Lambda$ case
\cite{MSY2,MSY3,MSSY5,MSSY6}.

However, we shall adopt a different set of quark distributions
from those used in Ref.~\cite{MSY4}. The reason is that the
next-to-leading order terms of every quark helicity distributions
were introduced in \cite{MSY4}, and this is meaningful only if
explicit data is available. The parameters still have large
freedom and there is actually less predictive power with such
parameters. Here our goal is different, since we want to
predict the rough features of the quark distributions, a similar
form with no adjustable parameters is enough for this purpose.
Therefore we only use the leading terms for quark helicity
distributions of the valence quarks
\begin{equation}
\begin{array}{cllr}
&q^{\uparrow}_{i}(x)=\frac{\tilde{A}_{q_{i}}}{B_3}
x^{-\frac{1}{2}}(1-x)^3;\\
&q^{\downarrow}_{i}(x)=\frac{\tilde{C}_{q_{i}}}{B_5}
x^{-\frac{1}{2}}(1-x)^5,
\end{array}
\label{case2}
\end{equation}
with  $i=1, 2$, where  $B_n=B(1/2,n+1)$ is the $\beta$-function
defined by $B(1-\alpha,n+1)=\int_0^1 x^{-\alpha}(1-x)^{n} {\mathrm
d} x$ for $\alpha=1/2$, and $B_3=32/35$ and $B_5=512/693$. From
(\ref{case2}),  we obtain  the valence quark normalization for
quark $q_i$
\begin{equation}
\begin{array}{clllc}
N_i=\tilde{A}_{q_{i}} +\tilde{C}_{q_{i}},
\end{array}
\label{Ni}
\end{equation}
and the corresponding polarized distribution in the
$J^p=\frac{1}{2}^+$ octet
\begin{equation}
\begin{array}{clllc}
\Delta Q_{i}&=&\tilde{A}_{q_{i}}  -\tilde{C}_{q_{i}},
\end{array}
\label{DQi}
\end{equation}
which can be adjusted from $\Sigma=\Delta u +\Delta d +\Delta s
\approx 0.3$, and the Bjorken sum rule
$\Gamma^p-\Gamma^n=\frac{1}{6}(\Delta u  -\Delta d ) =
\frac{1}{6} g_A/g_V \approx 0.2$, obtained in polarized DIS experiments \cite{Ma98}. The
coefficients $\tilde{A}_{q_{i}}$ and $\tilde{C}_{q_{i}}$ ($i=1,2$)
obtained in this way can make the ratio
\begin{equation}
R_A=\frac{\tilde{A}_{q_{1}}}{\tilde{A}_{q_{2}}}\label{RA}
\end{equation}
satisfy the $x \to 1 $ behavior of
$q^{\uparrow}_{1}(x)/q^{\uparrow}_{2}(x)$ for a baryon in the
SU(6) quark model. However, due to the non-collinearity of the
quarks, one cannot expect that the quark helicities will simply sum up
to the baryon spin, as the helicity distributions measured
on the light-cone are related by the Melosh-Wigner rotation to the
ordinary spins of the quarks in the quark model \cite{Ma91b}.
Furthermore, the coefficients for every baryon are completely
connected to each other by the SU(3) symmetry between the octet
baryons, which means
\begin{equation}
\begin{array}{lllc}
u^p=d^n=u^{\Sigma^+}=d^{\Sigma^-}=s^{\Xi^-}=s^{\Xi^0}
=\frac{2}{3}u^{\Lambda}+\frac{4}{3}s^{\Lambda}=2 u ^{\Sigma^0}=2 d
^{\Sigma^0};\\
d^p=u^n=s^{\Sigma^+}=s^{\Sigma^-}=d^{\Xi^-}=u^{\Xi^0}
=\frac{4}{3}u^{\Lambda}-\frac{1}{3}s^{\Lambda}=s ^{\Sigma^0}.
\end{array}
\end{equation}
Therefore there are actually no free parameters in this set of
pQCD based quark distributions for the whole set of octet baryons,
and the predictive ability of the approach is guaranteed. This set
of pQCD quark distributions corresponds to a revised version of
case 2 for the $\Lambda$ in Ref.~\cite{MSY3}. Of course, in this paper
we
are concerned by the $x \to 1$ behavior of the
valence quark distributions, so that the predictions should be
considered to be valid qualitatively rather than quantitatively,
and the model can be improved quantitatively by adding higher
order terms in the quark distributions \cite{MSY3,MSSY5}. The
parameters for quark distributions of octet baryons can be found
from Table 2.

\vspace{0.5cm}

\centerline{Table 2~~ The  parameters for quark distributions of
octet baryons  in pQCD}

\vspace{0.3cm}

\begin{footnotesize}
\begin{center}
\begin{tabular}{|c|c|c|c||c|c|c|c|c|c|}\hline
 Baryon & $q_1$ & $q_2$ & $R_A$ & $\Delta Q_1$ &
 $\Delta Q_2$ & $\tilde{A}_{q_1}$
 &$\tilde{C}_{q_1}$ &$\tilde{A}_{q_2}$
 &$\tilde{C}_{q_2}$ \\ \hline
p & u & d & 5 & 0.75 & -0.45 & 1.375 & 0.625 & 0.275 & 0.725 \\
\hline n & d & u & 5 & 0.75 & -0.45 & 1.375 & 0.625 & 0.275 &
0.725
\\ \hline $\Sigma^{+}$ & u & s & 5 & 0.75 & -0.45 & 1.375 & 0.625
& 0.275 & 0.725  \\ \hline $\Sigma^{0}$ & u(d) & s & $\frac{5}{2}$
& 0.375 & -0.45 & 0.6875 & 0.3125 & 0.275 & 0.725
\\ \hline $\Sigma^{-}$ & d  & s & 5 & 0.75 & -0.45 & 1.375
& 0.625 & 0.275 & 0.725  \\ \hline $\Lambda^{0}$ & s & u(d) & $ 2
$ &0.65 &-0.175 & 0.825 & 0.175 & 0.4125 & 0.5875
\\ \hline $\Xi^{-}$ & s & d & $5$ & 0.75 & -0.45 & 1.375 &
0.625 & 0.275 & 0.725
\\ \hline
$\Xi^{0}$ & s & u & $5$ & 0.75 & -0.45 & 1.375 & 0.625 & 0.275 &
0.725  \\ \hline

\end{tabular}
\end{center}
\end{footnotesize}

\vspace{0.5cm}

\section{Baryon polarizations in $e^+e^-$-annihilation at two energies}

In the standard model of electroweak interactions, the produced
quarks and antiquarks should be polarized in the unpolarized
$e^+e^-$-annihilation process due to the parity-violating coupling
of the fermions, and this leads to the polarizations of the
baryons (antibaryons) from the decays of the quarks (antiquarks).
Therefore we can study the polarized quark to hadron
fragmentations by the semi-inclusive production of hadrons in
$e^+e^-$-annihilation \cite{Gus93,Bur93,Bor98,Flo98b,MSY3,Liu00}.

The polarizations of the initial quarks from $e^+e^-$-annihilation
are given by
\begin{equation}
P_q=-\frac{A_q(1+\cos ^2 \theta )+B_q \cos \theta} {C_q(1+\cos ^2
\theta )+D_q \cos \theta},
\label{Pq}
\end{equation}
where
\begin{equation}
A_q=2 \chi_{2}(v_e^2+a_e^2)v_qa_q-2 e_q \chi_1 a_q v_e,
\end{equation}
\begin{equation}
B_q=4 \chi_2  v_e a_e (v_q^2+a_q^2)-4 e_q \chi_1 a_e v_q,
\end{equation}
\begin{equation}
C_q=e_q^2-2 \chi_1 v_e v_q e_q+ \chi_2 (a_e^2+v_e^2)
(a_q^2+v_q^2),
\end{equation}
\begin{equation}
D_q=8 \chi_2 a_e a_q v_e v_q-4 \chi_1 a_e a_q e_q,
\end{equation}
in which
\begin{equation}
\chi_1=\frac{1}{16 \sin^2 \theta_W \cos^2 \theta_W}
\frac{s(s-M_Z^2)}{(s-M_Z^2)^2+M_Z^2\Gamma_Z^2},
\end{equation}
\begin{equation}
\chi_2=\frac{1}{256 \sin^4 \theta_W \cos^4 \theta_W}
\frac{s^2}{(s-M_Z^2)^2+M_Z^2\Gamma_Z^2},
\end{equation}
\begin{equation}
a_e=-1
\end{equation}
\begin{equation}
v_e=-1+4 \sin^2 \theta_W
\end{equation}
\begin{equation}
a_q=2 T_{3q},
\end{equation}
\begin{equation}
v_q=2 T_{3q}-4 e_q \sin^2 \theta_W,
\end{equation}
where $T_{3q}=1/2$ for $u$, $c$, while $T_{3q}=-1/2$ for $d$, $s$,
$b$ quarks, $N_c=3$ is the color number, $e_q$ is the charge of
the quark in units of the proton charge, $\theta$ is the angle
between the outgoing quark and the incoming electron, $\theta_W$
is the Weinberg angle with $sin^2 \theta_W=0.2312$, and
$M_Z=91.187$~GeV and $\Gamma_Z=2.49$~GeV are the mass and width of
$Z^0$ \cite{PDG}. Averaging over $\theta$ for Eq.~(\ref{Pq}), one
obtains $P_q=-0.67$ for $q=u$, $c$, $P_q=-0.94$ for $q=d$, $s$,
and $b$, and $r_{d/u}=C_d/C_u=1.29$ at the $Z$-pole, i.e., the LEP
I energy $\sqrt{s}=91.187$~GeV. However, at the LEP II energy
$\sqrt{s}=200$~GeV, we can find that the $\gamma^*/Z^0$
interference should contribute since the terms involving $\chi_1$
do not vanish. One obtains, $P_q=-0.29$ for $q=u$, $c$,
$P_q=-0.71$ for $q=d$, $s$, and $b$, and $r_{d/u}=0.61$, which are
different from those at LEP I. We notice that the absolute value
of $P_q$ decreases with the energy square $s$ from LEP I to LEP II, 
as a result of non-zero $\chi_1$.

 From the cross section formulae \cite{Bur93,MSY3} for the
unpolarized and polarized hadron $h$ production, we can write the
formula for the hadron $h$ polarization
\begin{equation}
P_{h}(\theta)=-\frac{\sum\limits_{q} \left\{A_q (1+\cos^2
\theta)[\Delta D_q^h(z)-\Delta D_{\bar q}^h(z)] + B_q \cos \theta
[\Delta D_q^h(z)+\Delta D_{\bar q}^h(z)]\right\}}{\sum\limits_{q}
\left\{C_q (1+\cos^2 \theta)[D_q^h(z)+D_{\bar q}^h(z)] + D_q \cos
\theta [ D_q^h(z)+ D_{\bar q}^h(z)]\right\}}.
\end{equation}
By averaging over $\theta$ we obtain
\begin{equation}
P_{h}=-\frac{\sum\limits_{q} A_q [\Delta D_q^h(z)-\Delta D_{\bar
q}^h(z)]}{\sum\limits_{q} C_q [D_q^h(z)+D_{\bar q}^h(z)]}.
\label{PL2}
\end{equation}
We now present in Fig.~\ref{mssy7f1} the calculated hadron
polarizations for the octet baryons by adopting the Ansatz of
Eq.~(\ref{GLR}) connecting the fragmentation functions with the
quark distributions, in both the quark-diquark model and the pQCD
based analysis presented in the previous section.

\begin{figure}
\begin{center}
\leavevmode {\epsfysize=18cm \epsffile{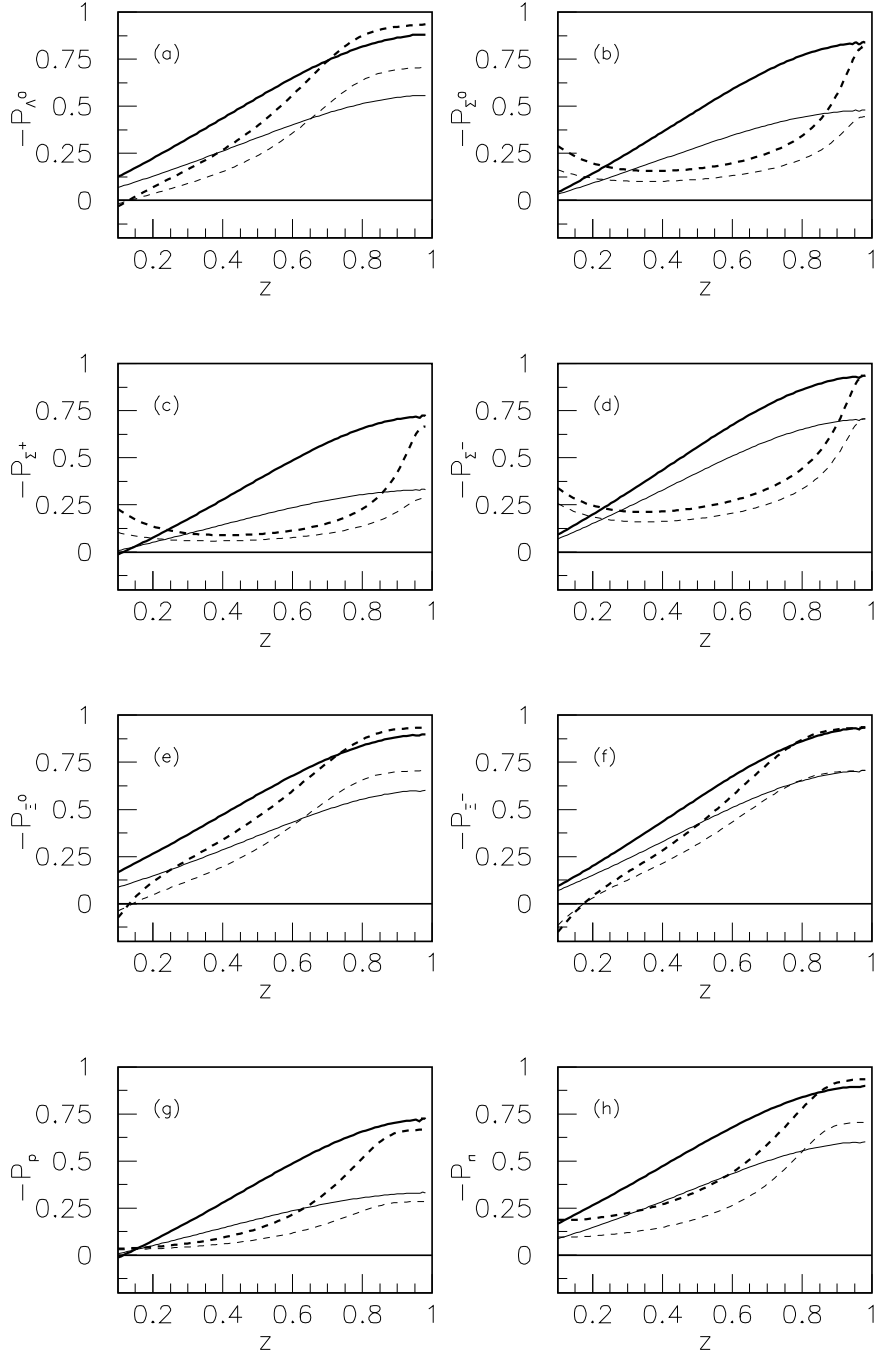}}
\end{center}
\caption[*]{\baselineskip 13pt The prediction of the longitudinal
hadron polarizations for the octet baryons in
$e^+e^-$-annihilation at two energies: LEP I at $Z$ resonance
$\sqrt{s} \approx 91$~GeV (thick curves) and LEP II at $\sqrt{s}
\approx 200$~GeV (thin curves), with input fragmentation functions
adopting the Ansatz Eq.~(\ref{GLR}) from valence quark
distributions in the pQCD based analysis (solid curves) and the
quark-diquark model (dashed curves). } \label{mssy7f1}
\end{figure}

There have been measurements of the $\Lambda$-polarization near
the $Z$-pole \cite{ALEPH96,DELPHI95,OPAL97}, i.e., at LEP I
energy, and we have already shown that the predictions of the two
models are compatible with the data \cite{MSY3}.
At LEP II energy, the predictions for the $\Lambda$ in the two
models are also  qualitatively similar, as can be seen from
Fig.~\ref{mssy7f1}(a). Therefore in order to distinguish between
the two models we need high precision measurements at large $z$.
For $\Xi^0$ and $\Xi^-$, the predictions for the polarizations in
the two models are also close to each other, and a clear
distinction between the two models is not easy, although there is
some difference between the $z \to 1$ behavior for $\Xi^0$ in both
models. However, we notice from Fig.~\ref{mssy7f1}(b)-(d) that the
predictions for $\Sigma$'s are quite different in the two models,
and it is possible to distinguish between them by the qualitative
behavior of the $\Sigma$ polarizations in the medium to large $z$
region. This can be easily understood since the $\Sigma$'s have
the most significant difference in the flavor and spin structure
between the two models in the medium to large $x$ region
\cite{MSY4}. It is also interesting that our predictions for the
$\Sigma$ polarizations are in an opposite direction to the
predictions by Liu and Liang, based on Monte-Carlo event
generators of two different pictures of quark polarizations inside
the $\Lambda$ without a physical mechanism for the $z$-dependence,
as can be seen by comparing our Fig.~\ref{mssy7f1}(c)-(d) with
their Fig.~10 \cite{Liu00}.

The different trends for the polarizations of $\Sigma$'s can be
easily understood. The contributions of $u$, $d$ and $s$ quarks to
the polarization of the produced hadron are of the same order as
can be seen from the calculated ratio $r_{d/u}$ in
$e^+e^-$-annihilation. In the pQCD based analysis, both the $s$
quark and the $u$ ($d$) quarks for $\Sigma^+$ ($\Sigma^-$) are
positively polarized inside the $\Sigma$'s at large $x$, therefore
the spin transfer from the initial quarks to the produced hadron
is positive. In the quark-diquark model, the $s$ quarks inside
$\Sigma$'s are negatively polarized, and their distribution is big
in the medium to large $x$ region; but the $u$ and $d$ quarks are
positively polarized and dominant at $x \to 1$, as can be seen
from Figs.~3-4 in Ref.~\cite{MSY4}, and therefore the contribution
from the $s$ quark causes a decrease in the spin transfer from all
initial quarks in the medium to large $z$ region. However, in the
analysis based on Monte-Carlo event generators \cite{Liu00}, the
$s$ quark contribution to the hadron fragmentation is dominant at
$z > 0.5$ so that the spin of the produced $\Sigma$'s comes from
the initial $s$ quarks, which are negatively polarized inside
$\Sigma$'s in their inputs, therefore the produced $\Sigma$'s
should be negatively polarized in comparison to the quark
polarization. We know that the $\Sigma^{\pm}$ polarizations can be
measured in experiments \cite{SigmaP} and the $\Sigma^{\pm}$
fragmentations can also be measured in $e^+e^-$-annihilation
\cite{SBaryon}. Therefore there should be no problem to measure
the $\Sigma^{\pm}$ in $e^+e^-$-annihilation with the available
experimental technique, and this can provide a sensitive test of
the above different predictions.

Nevertheless, in both the pQCD based analysis and the
quark-diquark model, the $s$ quarks inside $\Xi$'s are positively
polarized, and they contribute dominantly to the polarizations of
the produced $\Xi$'s at large $z$. This is also true in the
Monte-Carlo analysis, and we thus arrive at the same conclusion as
Liu and Liang \cite{Liu00} that there is little theoretical
uncertainty to predict the $\Xi$ polarizations at large $z$, as
can be seen from our Fig.~\ref{mssy7f1}(e)-(f) and Fig.~10 in
\cite{Liu00}. But there are still quantitative difference in the
predictions and high precision measurements can tell the
difference. This implies that the polarizations of $\Sigma$'s
produced in $e^+e^-$-annihilation can provide clearer
information to distinguish between the quark contributions to the
hadron fragmentation from different flavors than the $\Xi$'s,
therefore the $\Sigma$ polarizations in $e^+e^-$-annihilation
deserve experimental attention.

\section{Baryon polarizations in polarized
charged lepton DIS process}

We now look at the spin transfers for a hadron $h$ production in
polarized charged lepton DIS process. For a longitudinally
polarized charged lepton beam and an unpolarized nucleon target,
the longitudinal spin transfer to the fragmented hadron $h$ is
given in the quark parton model by \cite{Jaf96}
\begin{equation}
A^{h}(x,z)= \frac{\sum\limits_{q} e_q^2 [q^N(x,Q^2) \Delta
D_q^h(z,Q^2) + ( q \rightarrow \bar q)]}{\sum\limits_{q} e_q^2
[q^N (x,Q^2) D^h_q(z,Q^2) + ( q \rightarrow \bar q)]}~.
\label{DL}
\end{equation}
Here $y=\nu/E$, $x=Q^2/{2M_N \nu}$, and $z=E_h /\nu$, where
$q^2=-Q^2$ is the squared four-momentum transfer of the virtual
photon, $M_N$ is the proton mass, and $\nu$, $E$, and $E_{h}$ are
the energies of the virtual photon, the target nucleon, and the
produced hadron $h$ respectively, in the target rest frame;
$q^N(x,Q^2)$ is the quark distribution for the quark $q$ in the
target nucleon, $D_q^h (z,Q^2)$ is the fragmentation function for
$h$ production from quark  $q$, $\Delta D _q^h (z, Q^2) $ is the
corresponding longitudinal spin-dependent fragmentation function,
and $e_q$ is the quark charge in units of the elementary charge
$e$. For $\bar h $ production the spin transfer $A^{\bar h}(x,z)$
is obtained from Eq.~(\ref{DL}) by replacing hadron $h$ by
anti-hadron $\bar h$. The $h$ and $\bar h$ fragmentation functions
are related since we can safely assume matter-antimatter symmetry,
{\it i.e.}, $D^h_{q,\bar{q}}(z)=D^{\bar{h}}_{\bar{q},q}(z)$ and
similarly for $\Delta D^h_{q,\bar{q}}(z)$.

Recently, the HERMES Collaboration at DESY reported the  preliminary 
result of
the longitudinal spin transfer to the $\Lambda$ in polarized
positron DIS on the proton \cite{HERMES}. The E665 Collaboration
at FNAL also measured the $\Lambda$ and $\bar{\Lambda}$ spin
transfers from muon DIS \cite{E665}, and they observed very
different behaviour for $\Lambda$ and $\bar{\Lambda}$
polarizations, which might be related to the quark/antiquark
asymmetry of the nucleon sea either in the fragmentation functions
or in the distribution functions of the target \cite{MSSY5,MSSY6}.
The available data are consistent with both the quark-diquark
model and the pQCD based analysis for the quark to $\Lambda$
fragmentation functions \cite{MSY2,MSSY5,MSSY6}, and this again
supports the use of the Gribov-Lipatov relation as an Ansatz to
connect fragmentation functions with distribution functions. In
this section we only discuss the contribution from valence quarks
in the fragmentation functions, and therefore the predictions
should be only reasonable qualitatively at large $z$.

\begin{figure}
\begin{center}
\leavevmode {\epsfysize=18cm \epsffile{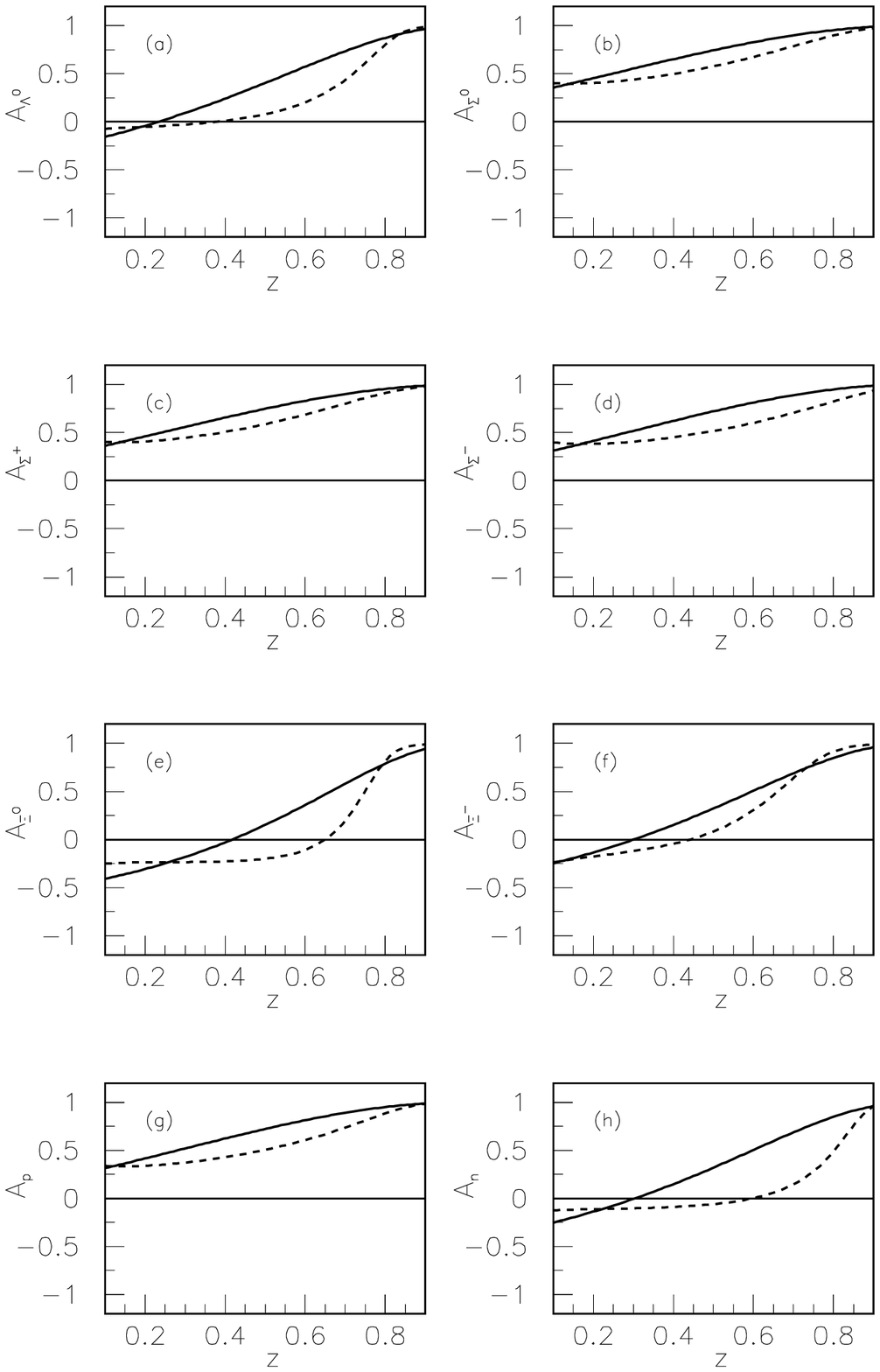}}
\end{center}
\caption[*]{\baselineskip 13pt The predictions of the
$z$-dependence for the $hadron$ spin transfer in polarized charged
lepton DIS process for the octet baryons. We adopt the CTEQ5 set 1
quark distributions \cite{CTEQ5} for the target proton at
$Q^2=2.5$~GeV$^2$ with the Bjorken variable $x$ integrated over
$0.02 \to 0.4$. The corresponding input fragmentation functions
adopt the Ansatz Eq.~(\ref{GLR}) from valence quark distributions
in the pQCD based analysis (solid curves) and the quark-diquark
model (dashed curves). }\label{mssy7f21}
\end{figure}

\begin{figure}
\begin{center}
\leavevmode {\epsfysize=18cm \epsffile{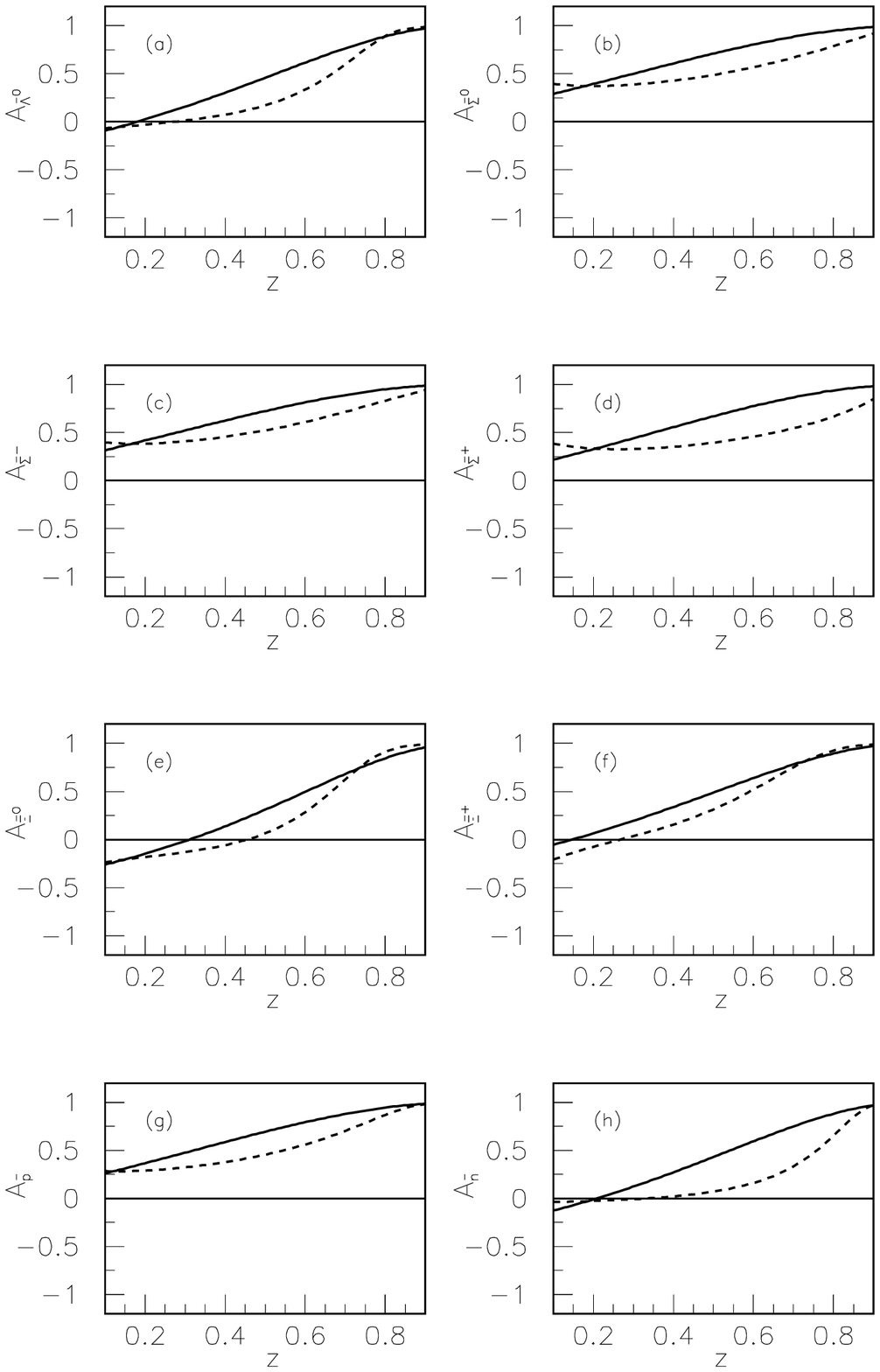}}
\end{center}
\caption[*]{\baselineskip 13pt The predictions of the
$z$-dependence for the $anti$-$hadron$ spin transfer in polarized
charged lepton DIS process for the octet anti-baryons. The others
are the same as Fig.~\ref{mssy7f21}. }\label{mssy7f22}
\end{figure}

We extend our analysis of the spin transfers for the octet baryons
and present our results in Figs.~\ref{mssy7f21}-\ref{mssy7f22},
where the spin transfers for baryons and anti-baryons are given
respectively. There is no much difference between the two figures
since the integrated $x$ range of the target quark distributions
is $0.02 \to 0.4$ where both quarks and antiquarks are important.
 From Fig.~\ref{mssy7f21} and Fig.~\ref{mssy7f22} we notice that
the predictions of the spin transfers are very similar between the
pQCD based analysis and the quark-diquark model for most baryons
and anti-baryons, except for $\Xi^0$ and $n$ where the spin
transfers might be negative at medium $z$ in the quark-diquark
model. This can be easily understood since the dominant
fragmentation chains contributing to this behavior are $u \to
\Xi^0$ and $u \to n$ as the $u$ quarks should be negatively
polarized at $x \to 1$ inside $\Xi^0$ and $n$, and the square
charge factor of the $u$ quarks is $4/9$ compared to $1/9$ for the
$d$ quarks. Also the larger number of $u$ quarks inside the
proton target amplifies this negative contribution. We propose to
measure the $\Xi^0$ transfer in the polarized DIS process to check
the two different predictions, as the experimental technique for
measuring $\Xi^0$ polarization is also well developed \cite{XiP}.

\section{Baryon polarizations in neutrino/antineutrino
DIS process}

One advantage of neutrino (antineutrino) process is that the
scattering of a neutrino beam on a hadronic target provides {\it a
source of polarized quarks with specific flavor structure}, and
this particular property makes the neutrino (antineutrino) process
an ideal laboratory to study the flavor-dependence of quark to
hadron fragmentation functions, especially in the polarized case
\cite{Ma99}. For the production of any hadron $h$ from neutrino
and antineutrino DIS processes, the longitudinal polarizations of
$h$ in its momentum direction, for $h$ in the current
fragmentation region can be expressed as \cite{MSSY5},
\begin{equation}
P_\nu^h(x,y,z)=-\frac{[d(x)+\varpi s(x)] \Delta D _u^h (z) -( 1-y)
^2 \bar{u} (x) [\Delta D _{\bar{d}}^h (z)+\varpi \Delta
D_{\bar{s}}^h(z)]} {[d(x)+\varpi s(x)] D_u ^h (z) + (1-y)^2
\bar{u} (x) [D _{\bar{d}}^h (z)+\varpi D_{\bar{s}}^h(z)]}~,
\label{nDIS}
\end{equation}

\begin{equation}
P_{\bar{\nu}}^h (x,y,z)=-\frac{( 1-y) ^2 u (x) [\Delta D _d^h
(z)+\varpi \Delta D _s^h (z)]-[\bar{d}(x)+\varpi \bar{s}(x)]
\Delta D _{\bar{u}}^h (z)}{(1-y)^2 u (x) [D _d^h (z)+\varpi D _s^h
(z)]+[\bar{d}(x)+\varpi \bar{s}(x)] D_{\bar{u}} ^h (z)}~,
\label{anDIS}
\end{equation}
where the terms with the factor $\varpi=\sin^2 \theta_c/\cos^2
\theta_c$ ($\theta_c$ is the Cabibbo angle) represent Cabibbo
suppressed contributions. We have neglected the charm
contributions both in the target and in hadron $h$. The detailed
$x$-, $y$-, and $z$- dependencies can provide more information
concerning the various fragmentation functions. As a special case,
the $y$-dependence can be simply removed by integrating over the
appropriate energy range and we can also integrate the
$x$-dependence to increase the statistics in experimental data
treatments.

In the $\Lambda$ case there is an interchange symmetry between the
$u$ and $d$ quarks: $u \leftrightarrow d$, which in general is not
present for other hadrons. After considering the symmetries
between different quark to hadron and anti-hadron fragmentation
functions \cite{Ma99}, there should be 8 independent fragmentation
functions which can be measured in neutrino (antineutrino) DIS
process for each hadron $h$,
\begin{equation}
D_u^{h}, ~~ D_{\bar{u}}^{h}, ~~ D_d^{h}+\varpi D_s^{h}, ~~
D_{\bar{d}}^{h} +\varpi D_{\bar{s}}^{h}, \label{uFF}
\end{equation}
and
\begin{equation}
\Delta D_u^{h}, ~~ \Delta D_{\bar{u}}^{h}, ~~ \Delta
D_d^{h}+\varpi \Delta D_s^{h}, ~~ \Delta D_{\bar{d}}^{h} +\varpi
\Delta D_{\bar{s}}^{h}. \label{pFF}
\end{equation}
Different combinations of unpolarized and polarized $h$ and
$\bar{h}$ productions in neutrino and antineutrino processes, and
choices of specific kinematics regions with different $x$, $y$,
and $z$, can measure the above fragmentation functions
efficiently. Unlike the $\Lambda$ case where the $u
\leftrightarrow d$ symmetry can be used \cite{Ma99,MSSY5}, it is
not possible to separate the $d$ and $s$ quark fragmentation
functions for any hadron $h$ due to the Cabibbo mixing. However,
in combination with the flavor dependence of fragmentation
functions in $e^+e^-$-annihilation and polarized charged lepton
DIS process, it should be possible to extract the various $d$ and
$s$ quark fragmentation functions, provided the accuracy of the
data is high enough. Another advantage of the neutrino
(antineutrino) processes is that the antiquark to hadron
fragmentation can also be conveniently extracted, and this can be
compared to specific predictions concerning the antiquark
polarizations inside baryons.

\begin{figure}
\begin{center}
\leavevmode {\epsfysize=8cm \epsffile{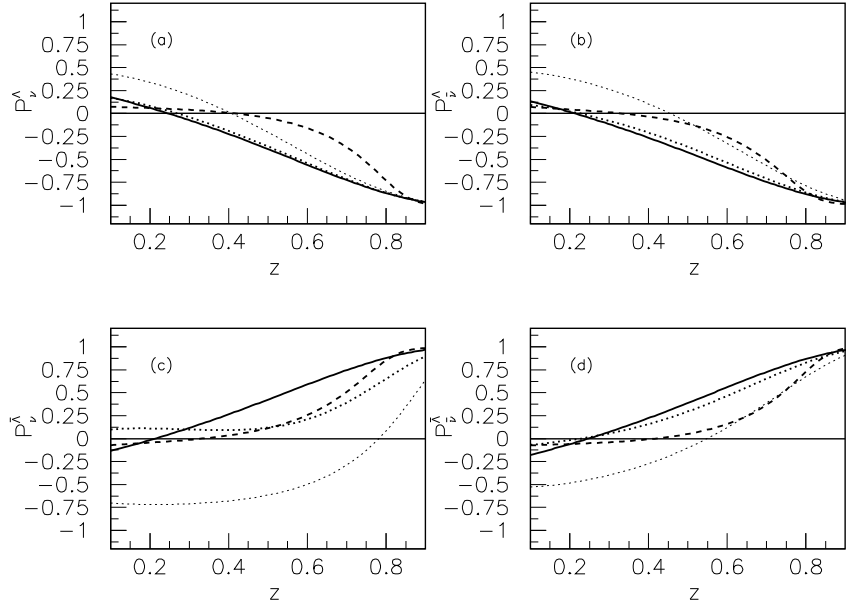}}
\end{center}
\caption[*]{\baselineskip 13pt The predictions of $z$-dependence
for the hadron and anti-hadron polarizations of $\Lambda$ in the
neutrino (antineutrino) DIS process. The solid and dashed curves
are results with input fragmentation functions adopting the Ansatz
Eq.~(\ref{GLR}) from valence quark distributions in the pQCD based
analysis (solid curves) and the quark-diquark model (dashed
curves), and the dotted curves are the results for the pQCD based
analysis of valence quarks with scenario I of asymmetric
quark-antiquark sea (thin dotted curves) and scenario II of
symmetric quark-antiquark sea (think dotted curves) in
Ref.~\cite{MSSY5}. We adopt the CTEQ5 set 1 quark distributions
\cite{CTEQ5} for the target proton at $Q^2=2.5$~GeV$^2$ with the
Bjorken variable $x$ integrated over $0.02 \to 0.4$ and $y$
integrated over $0 \to 1$. }\label{mssy7f31}
\end{figure}

\begin{figure}
\begin{center}
\leavevmode {\epsfysize=8cm \epsffile{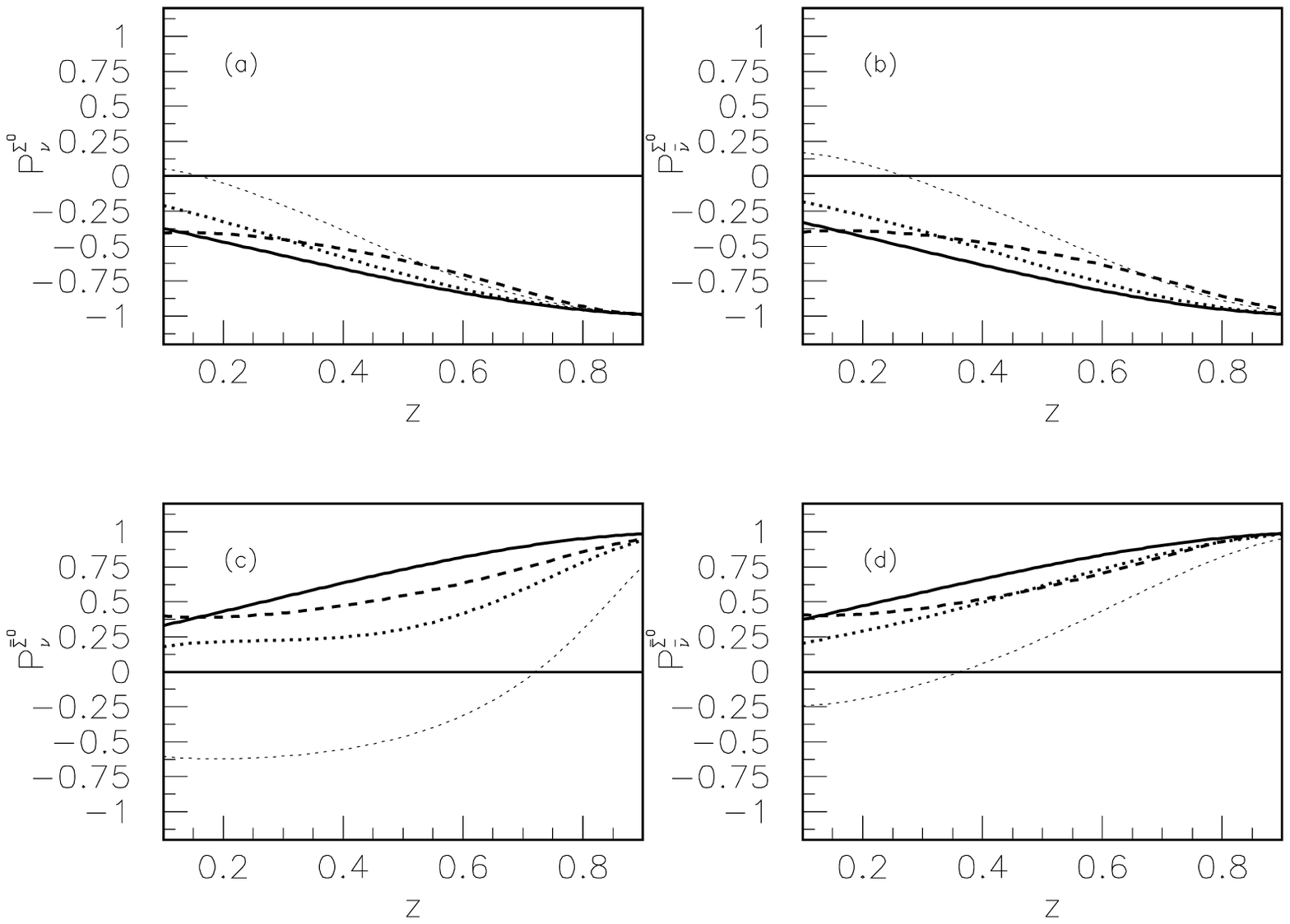}}
\end{center}
\caption[*]{\baselineskip 13pt The same as Fig.~\ref{mssy7f31},
but for predictions of $z$-dependence for the hadron and
anti-hadron polarizations of $\Sigma^0$ in the neutrino
(antineutrino) DIS process. }\label{mssy7f32}
\end{figure}

\begin{figure}
\begin{center}
\leavevmode {\epsfysize=8cm \epsffile{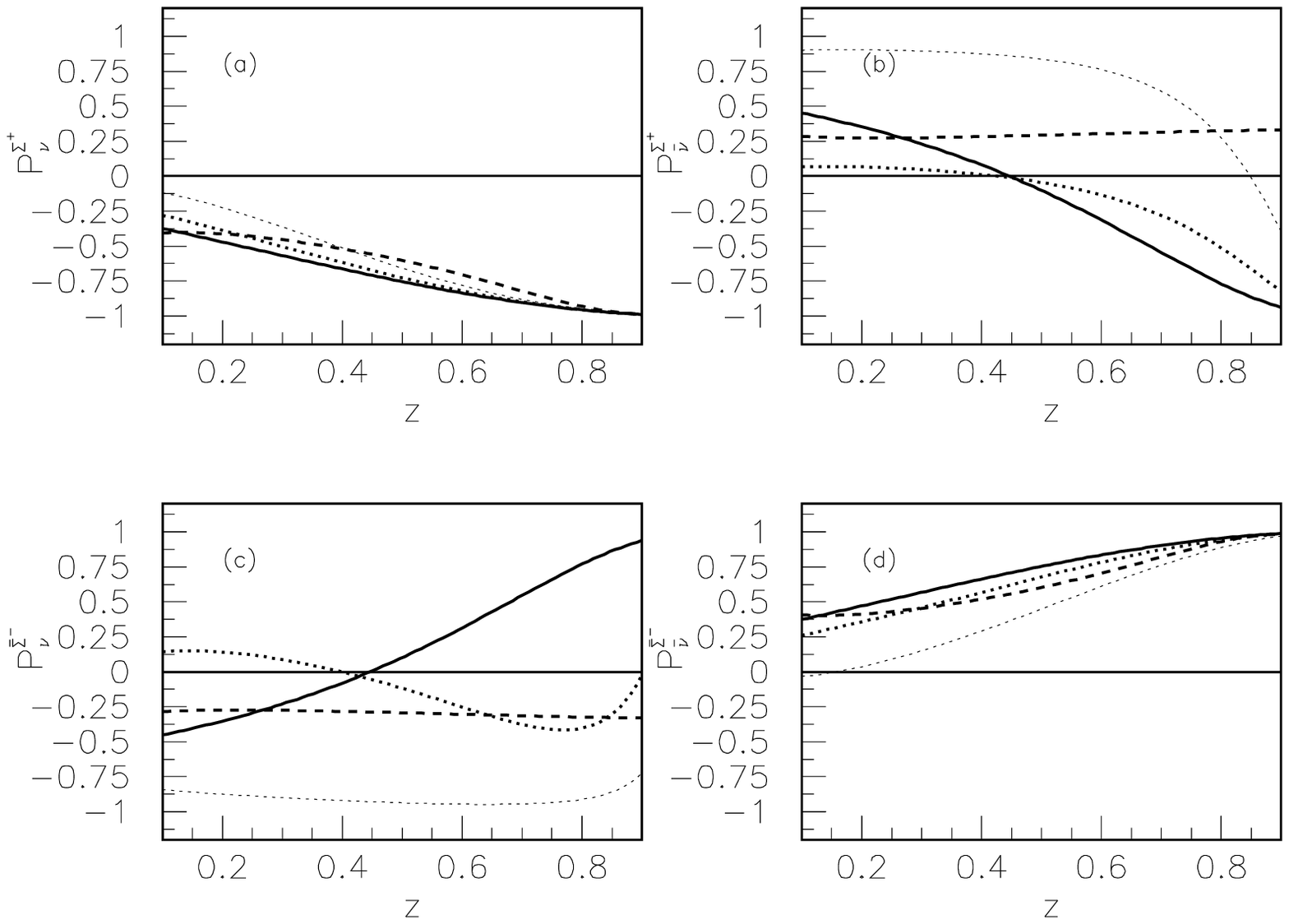}}
\end{center}
\caption[*]{\baselineskip 13pt The same as Fig.~\ref{mssy7f31},
but for predictions of $z$-dependence for the hadron and
anti-hadron polarizations of $\Sigma^+$ in the neutrino
(antineutrino) DIS process. }\label{mssy7f33}
\end{figure}

\begin{figure}
\begin{center}
\leavevmode {\epsfysize=8cm \epsffile{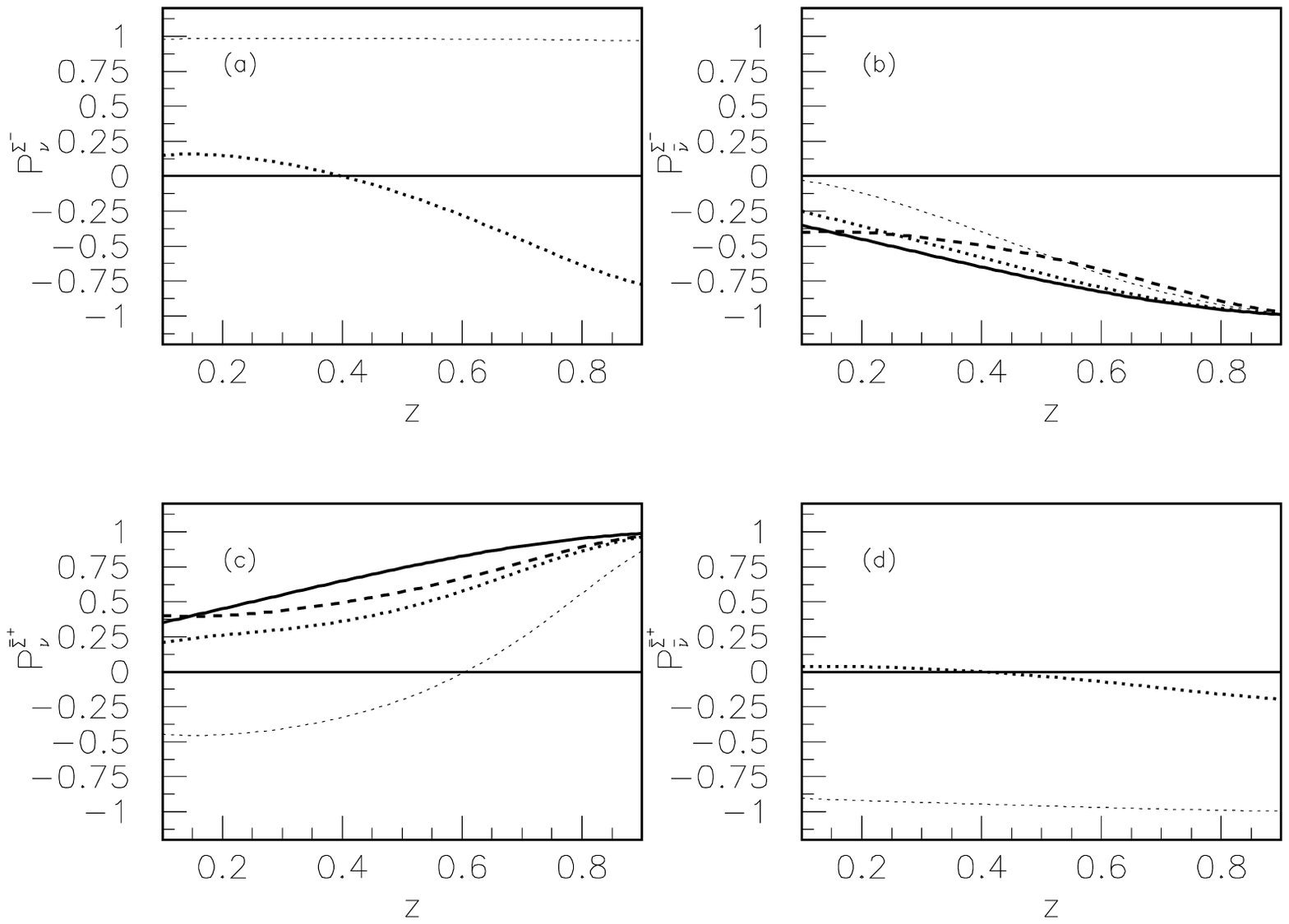}}
\end{center}
\caption[*]{\baselineskip 13pt The same as Fig.~\ref{mssy7f31},
but for predictions of $z$-dependence for the hadron and
anti-hadron polarizations of $\Sigma^-$ in the neutrino
(antineutrino) DIS process. }\label{mssy7f34}
\end{figure}

\begin{figure}
\begin{center}
\leavevmode {\epsfysize=8cm \epsffile{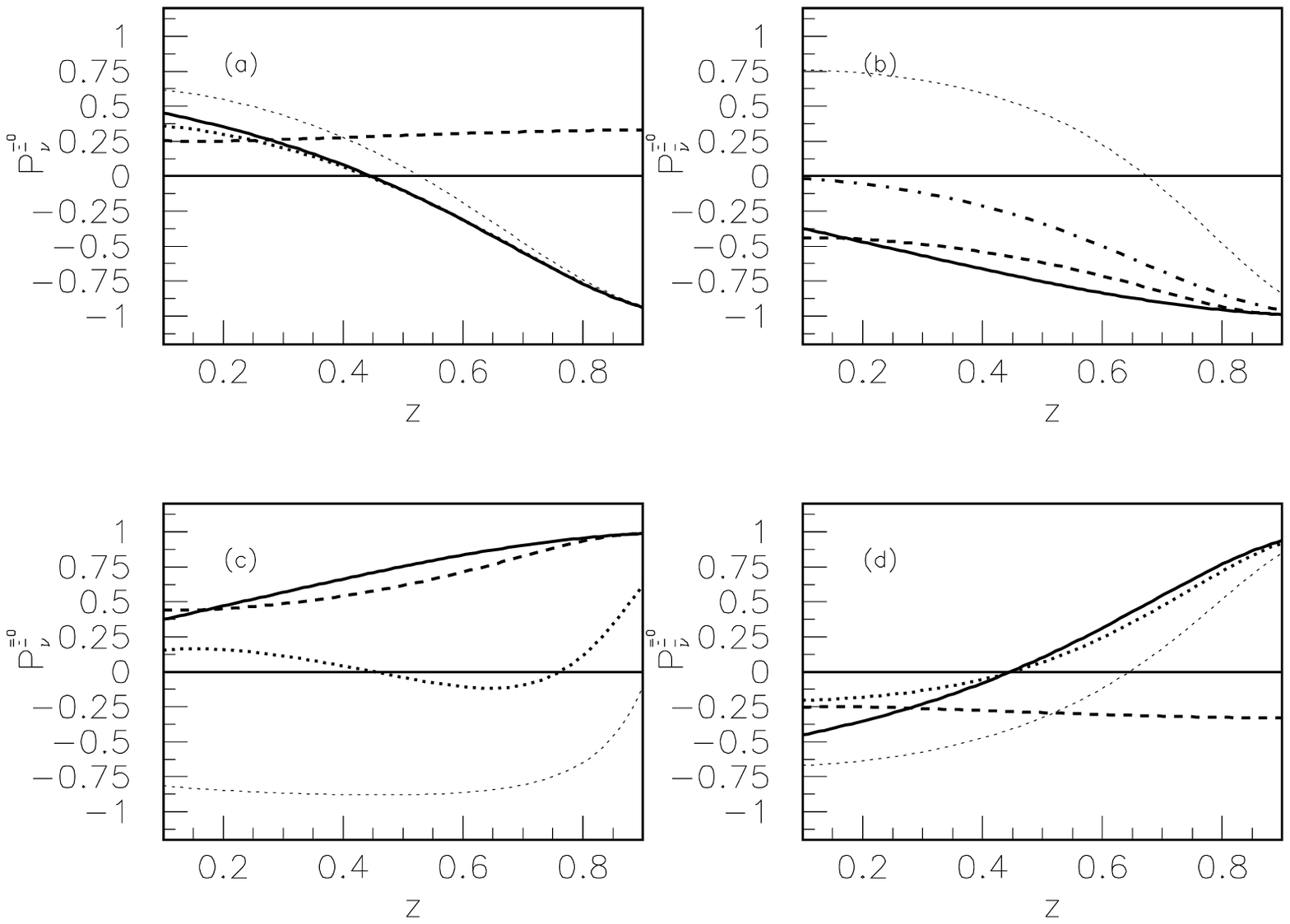}}
\end{center}
\caption[*]{\baselineskip 13pt The same as Fig.~\ref{mssy7f31},
but for predictions of $z$-dependence for the hadron and
anti-hadron polarizations of $\Xi^0$ in the neutrino
(antineutrino) DIS process. }\label{mssy7f35}
\end{figure}

\begin{figure}
\begin{center}
\leavevmode {\epsfysize=8cm \epsffile{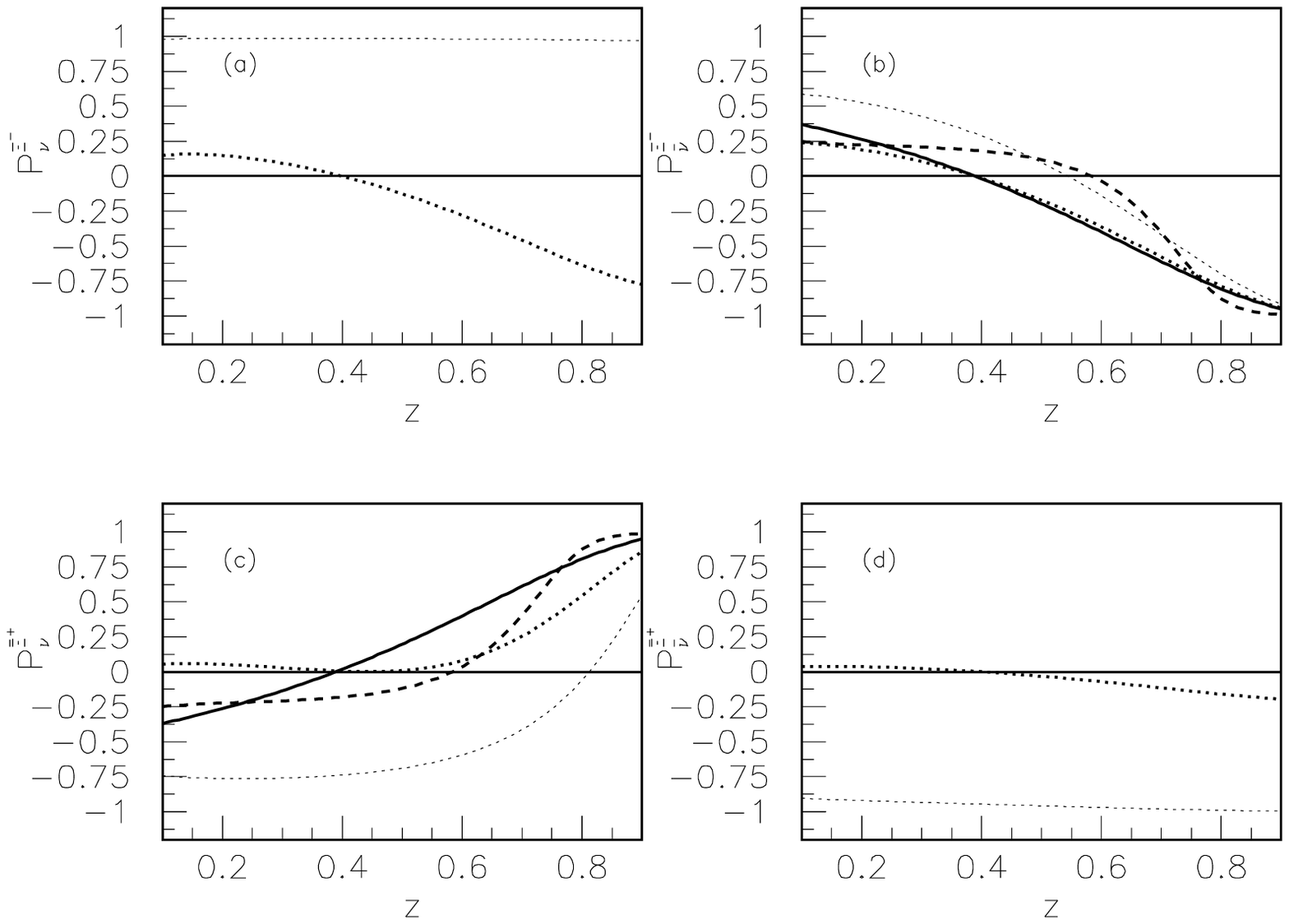}}
\end{center}
\caption[*]{\baselineskip 13pt The same as Fig.~\ref{mssy7f31},
but for predictions of $z$-dependence for the hadron and
anti-hadron polarizations of $\Xi^-$ in the neutrino
(antineutrino) DIS process. }\label{mssy7f36}
\end{figure}

\begin{figure}
\begin{center}
\leavevmode {\epsfysize=8cm \epsffile{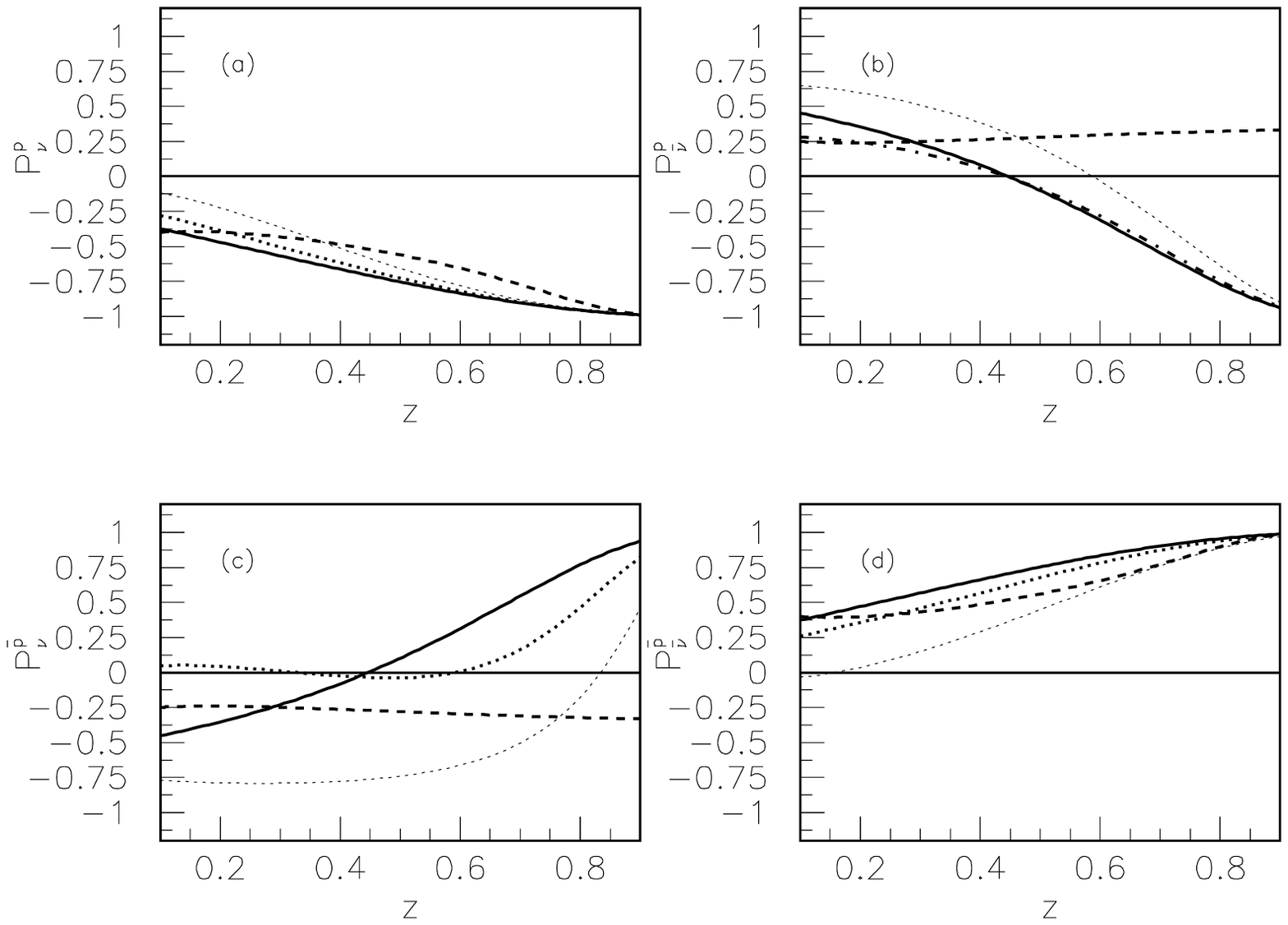}}
\end{center}
\caption[*]{\baselineskip 13pt The same as Fig.~\ref{mssy7f31},
but for predictions of $z$-dependence for the hadron and
anti-hadron polarizations of $p$ in the neutrino (antineutrino)
DIS process. }\label{mssy7f37}
\end{figure}

\begin{figure}
\begin{center}
\leavevmode {\epsfysize=8cm \epsffile{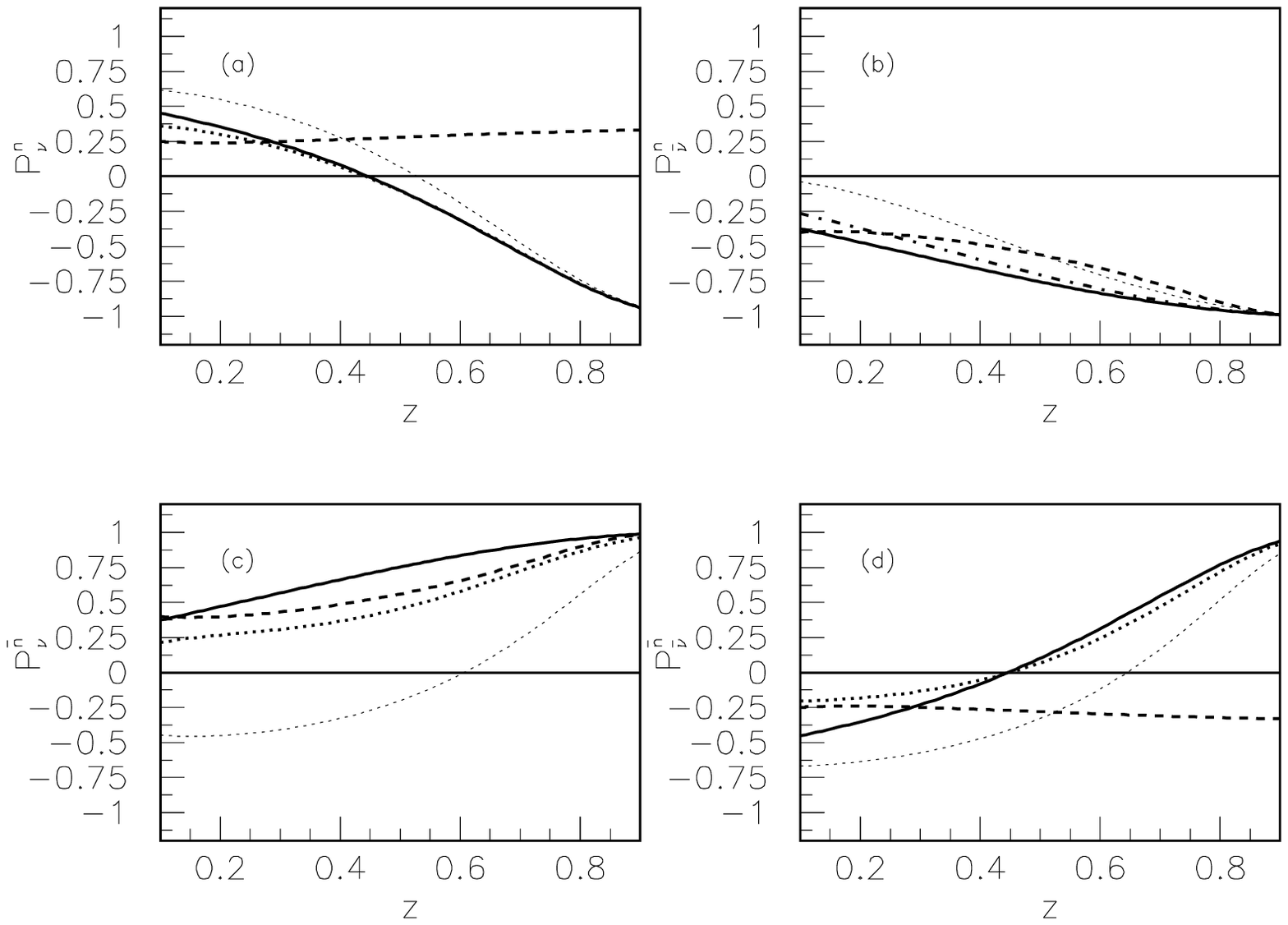}}
\end{center}
\caption[*]{\baselineskip 13pt The same as Fig.~\ref{mssy7f31},
but for predictions of $z$-dependence for the hadron and
anti-hadron polarizations of $n$ in the neutrino (antineutrino)
DIS process. }\label{mssy7f38}
\end{figure}

In Figs.~\ref{mssy7f31}-\ref{mssy7f38} we present our predictions
for the hadron and anti-hadron polarizations of the octet baryons
in neutrino and anti-neutrino DIS processes. Some precision data
on $\Lambda$ and $\bar{\Lambda}$ production have been taken by the
NOMAD neutrino beam experiment \cite{NOMAD}, and our predictions
for the $\Lambda$ polarization in both the pQCD based analysis and
the quark-diquark model, as presented in Fig.~\ref{mssy7f31}(a),
has been proved to be supported by the preliminary data. Notice
that another work \cite{Kot98} predicted quite different $\Lambda$
polarization compared to ours. Thus our knowledge of the quark to
$\Lambda$ fragmentation functions is improving. In principle, the
polarizations for $\Sigma^{\pm}$, $\Xi^0$, and $\Xi^-$ and their
anti-hadron partners can be also measured by the NOMAD
collaboration, therefore we can systematically study the flavor
decomposition of various quark to octet baryon fragmentation
functions. This will enrich our knowledge on the fragmentation
functions and the detailed quark structure of octet baryons.

Now we look only at the contributions from the valence quarks,
shown in Figs.~\ref{mssy7f31}-\ref{mssy7f38} as solid curves for
the pQCD based analysis and dashed curves for the quark-diquark
model. We find that the $\Sigma^+$ and $\Xi^0$ have significant
different predictions between the two models, as can be seen from
Fig.~\ref{mssy7f33}(b) and (c) for $\Sigma^+$, and from
Fig.~\ref{mssy7f35}(a) and (d) for $\Xi^0$. The difference can be
understood as follows. For the $\Sigma^+$ production in
$\bar{\nu}$ DIS, we can see from Eq.~(\ref{anDIS}) that only the
$d$ and $s$ quarks contribute, whereas inside $\Sigma^+$ only the
$s$ valence quark contributes. The $s$ quark is positive polarized
at large $x$ in the pQCD based analysis, whereas it is negatively
polarized in the quark-diquark model. This gives opposite trends
of the $\Sigma^+$ polarizations at large $z$ in the two models.
This discussion can also be extended to $\bar{\Sigma}^-$ in $\nu$
DIS as shown in Fig.~\ref{mssy7f33}(c). For the $\Xi^0$ production
in $\nu$ DIS, as can be seen from Eq.~(\ref{nDIS}), only the $u$
quarks contribute and the $u$ valence quarks inside $\Xi^0$ are
positively polarized in the pQCD based analysis at large $x$
whereas it is negatively polarized in the quark-diquark model.
This gives different trends of the $\Xi^0$ polarizations at large
$z$ between the two models, and a similar discussion applies to
$\bar{\Xi}^0$ production in $\bar{\nu}$ DIS as shown in
Fig.~\ref{mssy7f35}(d). Thus the $\Sigma^+$ and $\Xi^0$
polarizations in neutrino (antineutrino) DIS process can test 
different predictions between the pQCD based analysis and the
quark-diquark model.

We now look at Fig.~\ref{mssy7f37}(b),(c) and
Fig.~\ref{mssy7f38}(a), (d), and find that a similar discussion
also applies to the nucleon. This means that measuring the nucleon
polarizations in the neutrino (antineutrino) DIS process, for
example $p$ polarization in $\bar \nu$ DIS, can provide a good
test for the different predictions of $\Delta d/d=-1/3$ at large
$x$ in the quark-diquark model \cite{Ma96} and $\Delta d/d=1$ at
$x \to 1$ in the pQCD based analysis \cite{Bro95}.

We also notice that there are no valence quark contributions to
$\Sigma^-$ and $\Xi^-$ productions in $\nu$ DIS process, and
similarly for the productions of their anti-baryon partners,
$\bar{\Sigma}^+$ and $\bar{\Xi}^+$, in $\bar{\nu}$ DIS process.
This means that the above channels are the most suitable in order
to study the contributions from antiquarks to the fragmentation
functions. In order to show the sensitivity to the sea quark
content of the octet baryons, we adopt the two scenarios of sea
quark distributions in Ref.~\cite{MSSY5} for the $\Lambda$ and
simply assume that the sea is the same for all octet baryons. This
assumption is surely not correct, but our knowledge of the sea
quark distributions inside hyperons is rather poor at the moment.
 From another point of view, the Gribov-Lipatov relation
Eq.~(\ref{GLR}) should be only valid at large $z$, and we should
consider our method as a phenomenological way to parameterize the
quark (antiquark) to baryon fragmentations in the small $z$
region. Therefore we should not rely
strongly on the predictions for the sea
quark content of the hyperons, and our
purpose is only to show that the results are rather sensitive to
different scenarios of the sea quarks. Scenario I of the sea
quarks corresponds to asymmetric quark-antiquark helicity
distributions, and scenario II corresponds to a symmetric case,
and the predictions including the sea contributions together with
the valence quarks in the pQCD based analysis are shown as dotted
curves (thin curves for scenario I and thick curves for scenario
II) in Figs.~\ref{mssy7f31}-\ref{mssy7f38}. From
Figs.~\ref{mssy7f34}(a) and \ref{mssy7f36}(a) we find that the
$\Sigma^-$ and $\Xi^-$ polarizations are rather sensitive to the
different scenarios and therefore a measure of these polarizations
in $\nu$ DIS process can provide important information concerning
the contributions from antiquark to hyperon fragmentations. We also
notice that the sea contributions also play an important role in
hyperon productions of neutrino (antineutrino) DIS process, as can
be seen from Figs.~\ref{mssy7f31}-\ref{mssy7f38}. Thus neutrino
(antineutrino) DIS process is a sensitive place to study the sea
quark content of hyperons, although this requires high precision
data.

\section{Summary and Conclusion}

In this paper we systematically extend our study of the spin and
flavor structure of the $\Lambda$, to all other hyperons of the
octet baryons, by considering quark fragmentation in three
different processes. The predictions of the $\Lambda$ longitudinal
polarizations in both the quark-diquark model and pQCD based
analysis, have been proved to be supported by the available data
for $\Lambda$ production in the three processes:
$e^+e^-$-annihilation, polarized charged lepton DIS, and neutrino
(antineutrino) DIS. We presented in this paper the predictions of
the hyperon longitudinal polarizations for the octet baryons
obtained by fragmentation in the above three processes, and
suggested sensitive tests to check the different predictions. We
find that the polarization of $\Sigma$ hyperons in $e^+e^-$
annihilation provides a new direction to test different
predictions. The $\Xi^0$ polarization in polarized charged lepton
DIS process on the proton target can also test the different
predictions between the pQCD based analysis and the quark-diquark
model. The $\Sigma^+$ and $\Xi^0$ polarizations in neutrino
(antineutrino) DIS process can test the different predictions
concerning the valence structure of the hyperons, whereas
$\Sigma^-$ and $\Xi^-$ polarizations are suitable to study the
antiquark to hyperon fragmentations. The predictions of this paper
are supposed to be valid qualitatively rather than quantitatively,
and the difference in the predictions can be understood by clear
physical pictures which can be tested explicitly by various
methods. We expect that systematic studies on the various hyperon
fragmentations, both theoretical and experimental, will enrich our
knowledge of the quark to hadron fragmentations and of the quark
structure of the octet baryons.

{\bf Acknowledgments: } This work is partially supported by
National Natural Science Foundation of China under Grant Numbers
19975052 and 19875024, by Fondecyt (Chile) postdoctoral fellowship
3990048, by the cooperation programmes Ecos-Conicyt and CNRS-
Conicyt between France and Chile, and by Fondecyt (Chile) grant
1990806 and 8000017, and by a C\'atedra Presidencial (Chile).


\newpage
\begin{thebibliography}{99}

\bibitem{Gus93}
G.~Gustafson and J.~H\"akkinen, Phys. Lett. {\bf B 303},
350(1993).

\bibitem{Bur93}
M. Burkardt and R.L. Jaffe, Phys. Rev. Lett. {\bf 70}, 2537
(1993).

\bibitem{Lu95}
W.~Lu and B.-Q.~Ma, Phys. Lett. {\bf B 357}, 419 (1995);

W.~Lu, Phys. Lett. {\bf B 373}, 223 (1996);

J.~Ellis, D.~Kharzeev, and A.~Kotzinian, Z. Phys. {\bf C 69}, 467
(1996).

\bibitem{Jaf96}
R.L.~Jaffe, Phys. Rev. {\bf D 54}, R6581 (1996).

\bibitem{Kot98}
A.~Kotzinian, A.~Bravar, and D.von Harrach, Eur. Phys. J. {\bf C
2}, 329 (1998).

\bibitem{Flo98}
D.de Florian, M.~Stratmann, and W.~Vogelsang, Phys. Rev. Lett.
{\bf 81}, 530 (1998).

\bibitem{Bor98}
C.~Boros and Z.~Liang, Phys. Rev. {\bf D 57}, 4491 (1998).

\bibitem{Flo98b}
D.de Florian, M.~Stratmann, and W.~Vogelsang, Phys. Rev. {\bf D
57}, 5811 (1998).

\bibitem{Ma99}
B.-Q. Ma and J. Soffer, Phys. Rev. Lett. {\bf 82}, 2250 (1999).

\bibitem{MSY2}
B.-Q. Ma, I. Schmidt, and J.-J. Yang, Phys. Lett. {\bf B 477}, 107
(2000).

\bibitem{MSY3}
B.-Q. Ma, I. Schmidt, and J.-J. Yang, Phys. Rev. {\bf D 61},
034017 (2000).

\bibitem{MSY4}
B.-Q. Ma, I. Schmidt, and J.-J. Yang, Nucl. Phys. {\bf B 574}, 331
(2000).

\bibitem{Nza95}
M.~Nzar and P.~Hoodbhoy, Phys.~Rev.~{\bf D 51}, 32 (1995).

\bibitem{Bor99b}
C.~Boros and and A.W.~Thomas, Phys.~Rev.~{\bf D 60}, 074017
(1999); C.~Boros, T.~Londergan, and A.W.~Thomas, Phys.~Rev.~{\bf D
61}, 014007 (2000); Phys. Rev. {\bf D 62}, 014021 (2000).

\bibitem{Ash99}
D.~Ashery and H.J.~Lipkin, Phys. Lett. {\bf B 469}, 263 (1999).

M. Anselmino, M. Boglione, and F. Murgia, 
Phys. Lett. {\bf B 481}, 253 (2000). 

\bibitem{MSSY5}
B.-Q. Ma, I. Schmidt, J. Soffer, and J.-J. Yang, 
Eur. Phys. J. {\bf C 16}, 657 (2000).

\bibitem{MSSY6}
B.-Q. Ma, I. Schmidt, J. Soffer, and J.-J. Yang, hep-ph/0005210,
Phys. Lett. {\bf B 488}, 254 (2000) in press.

\bibitem{Liu00}
C.~Liu and Z.~Liang, 
Phys. Rev. {\bf D 62}, 094001 (2000). 

\bibitem{ALEPH96}
ALEPH Collaboration, D. Buskulic {\it et al}, Phys. Lett. {\bf B
374}, 319 (1996).

\bibitem{DELPHI95}
DELPHI Collaboration, Report No.DELPHI 95-86 PHYS 521,
CERN-PPE-95-172, presented at the EPS-HEP 95 conference, Brussels,
1995.

\bibitem{OPAL97}
OPAL Collaboration, K. Ackerstaff {\it et al},
Eur. Phys. J. {\bf C 2}, 49 (1998).


\bibitem{HERMES}
HERMES Collaboration, A. Airapetian {\it et al.}, hep-ex/9911017.

\bibitem{E665}
E665 Collaboration, M. R. Adams {\it et al.}, hep-ex/9911004.

\bibitem{NOMAD}
NOMAD Collaboration, C. Lachaud, Th\`ese de Doctorat Univ. Denis
Diderot, Paris, 3/5/2000.

\bibitem{countingr}
R. Blankenbecler and S.J. Brodsky, Phys. Rev. {\bf D 10}, 2973
(1974); J.F. Gunion, Phys. Rev. {\bf D 10}, 242 (1974); S.J.
Brodsky and G.P. Lepage, in Proc. 1979 Summer Inst. on Particle
Physics, SLAC (1979).

\bibitem{Bro95}
S.J. Brodsky, M. Burkardt, and I. Schmidt, Nucl. Phys. {\bf B
441}, 197 (1995).

\bibitem{Ma96}
B.-Q. Ma, Phys. Lett. {\bf B 375}, 320 (1996).


\bibitem{GLR}
V.N.~Gribov and L.N.~Lipatov, Phys. Lett. {\bf B 37}, 78 (1971);
Sov. J. Nucl. Phys. {\bf 15}, 675 (1972).

\bibitem{Bro97}
S.J.~Brodsky and B.-Q.~Ma, Phys. Lett. {\bf B 392}, 452 (1997).

\bibitem{SigmaP}
Y.W.~Wah {\it et al}, Phys. Rev. Lett. {\bf 55}, 2551 (1985);
C.~Wilkinson {\it et al}, Phys. Rev. Lett. {\bf 58}, 855 (1987);
E761 Collaboration, A.~Morelos {\it et al.}, Phys. Rev. Lett. {\bf
71}, 2172 (1993).

\bibitem{XiP}
K.~Heller {\it et al.}, Phys. Rev. Lett. {\bf 51}, 2025 (1983);
P.M.~Ho {\it et al.}, Phys. Rev. Lett. {\bf 65}, 1713 (1990); J.
Duryea {\it et al.}, Phys. Rev. Lett. {\bf 67}, 1193 (1991).

\bibitem{Sigma0P}
B.E.~Bonner {\it et al.}, Phys. Rev. Lett. {\bf 62}, 1591 (1989).

\bibitem{Fey72}

R.P. Feynman, {\it Photon Hadron Interactions} (Benjamin, New
York, 1972), p. 150.


\bibitem{DQM}
F.E.~Close, Phys. Lett. {\bf 43 B}, 422 (1973); Nucl. Phys. {\bf B
80}, 269 (1974);

R.~Carlitz, Phys.~Lett.~{\bf B 58}, 345 (1975);

J.~Kaur, Nucl.~Phys.~{\bf B 128}, 219 (1977);

A.~Sch\"afer, Phys.~Lett.~{\bf B 208}, 175 (1988);

F.E.~Close and A.W.~Thomas, Phys. Lett. {\bf B 212}, 227 (1988);

N.~Isgur, Phys. Rev. {\bf D 59}, 034013 (1999).


\bibitem{Ma91b}
B.-Q.~Ma, J. Phys. {\bf G 17}, L53 (1991);

B.-Q.~Ma and Q.-R.~Zhang, Z.~Phys. {\bf C 58}, 479 (1993).

\bibitem{Ma98}
B.-Q. Ma, I. Schmidt, and J. Soffer, Phys. Lett. {\bf B 441}, 461
(1998);

B.-Q. Ma and I. Schmidt, Phys. Rev. {\bf D 58}, 096008 (1998).

\bibitem{BHL}
S. J. Brodsky, T. Huang, and G. P. Lepage, in {\it Particles and
Fields-2}, Proceedings of the Banff Summer Institute, Banff,
Alberta, 1981, edited by A. Z. Capri and A. N. Kamal (Plenum, New
York,1983), p. 143.

\bibitem{Hua94}
T. Huang, B.-Q. Ma, and Q.-X. Shen, Phys. Rev. {\bf D 49}, 1490
(1994).


\bibitem{Far75}
G.R. Farrar and D.R. Jackson, Phys. Rev. Lett. {\bf 35}, 1416
(1975).

\bibitem{Yang99}
U.K. Yang and A. Bodek, Phys. Rev. Lett. {\bf 82}, 2467 (1999).

\bibitem{Sch00}
I.~Schmidt and J.-J.~Yang, hep-ph/0005054.

\bibitem{PDG}
Particle Data Group, C.~Caso {\it et al.}, Euro. Phys. J. {\bf C
3}, 1 (1998).

\bibitem{SBaryon}
S. Abachi {\it et al.}, Phys. Rev. Lett. {\bf 58}, 2627 (1987).


\bibitem{CTEQ5}
CTEQ Collaboration, H. L. Lai et al., Eur. Phys. J. {\bf C 12},
375 (2000).

\nonfrenchspacing
\end{thebibliography}
\end{document}